\documentclass[aps,pra,superscriptaddress,showpacs,floatfix,longbibliography,notitlepage]{revtex4-2}
\usepackage{dsfont}
\usepackage{cprotect}
\usepackage{graphicx}
\usepackage{amsmath}
\usepackage{commath}
\usepackage{comment} 
\usepackage{amssymb}
\usepackage{hyperref}
\usepackage{bm}
\usepackage{subfigure}
\hypersetup{colorlinks=true,linkcolor=cyan,citecolor=magenta,filecolor=magenta,urlcolor=cyan,runcolor=cyan}
\graphicspath{{figures/}}
\usepackage[pdftex]{color}
\usepackage{algorithm}
\usepackage{algpseudocode}
\newcommand{\e}{\text{e}}
\newcommand{\beq}{\begin{equation}}
\newcommand{\eeq}{\end{equation}}
\newcommand{\beqnn}{\begin{equation*}}
\newcommand{\eeqnn}{\end{equation*}}
\newcommand{\bea}{\begin{eqnarray}}
\newcommand{\eea}{\end{eqnarray}}
\newcommand{\beann}{\begin{eqnarray*}}
\newcommand{\eeann}{\end{eqnarray*}}
\newcommand{\bes} {\begin{subequations}}
\newcommand{\ees} {\end{subequations}}

\newcommand{\Mod}[1]{\ (\mathrm{mod}\ #1)}
\newcommand\ini{\alpha}
\newcommand\fin{\omega}
\newcommand\algspace{\hspace{\algorithmicindent}}

\begin{document}
\title{Calculating Elements of Matrix Functions using Divided Differences}

\author{Lev Barash}
\affiliation{Landau Institute for Theoretical Physics, 142432 Chernogolovka, Russia}
\author{Stefan G\"uttel}
\affiliation{Department of Mathematics, The University of Manchester, M13\,9PL Manchester, United Kingdom}
\author{Itay Hen}
\email{itayhen@isi.edu}
\affiliation{Information Sciences Institute, University of Southern California, Marina del Rey, California 90292, USA}
\affiliation{Department of Physics and Astronomy and Center for Quantum Information Science \& Technology, University of Southern California, Los Angeles, California 90089, USA}

\begin{abstract}
\noindent We introduce a method for calculating individual elements of matrix functions. Our technique makes use of a novel series expansion for the action of matrix functions on basis vectors that is memory efficient even for very large matrices. We showcase our approach by calculating the matrix elements of the exponential of a transverse-field Ising model and evaluating quantum transition amplitudes for large many-body Hamiltonians of sizes up to $2^{64} \times 2^{64}$ on a single workstation. We also discuss the application of the method to matrix inverses. 
We relate and compare our method to the  state-of-the-art and demonstrate its advantages. We also discuss practical applications of our method.
\end{abstract}
\maketitle

\section{Introduction}

The evaluation of functions of matrices (or matrix functions) plays an important role in many scientific applications, ranging from differential equations to nuclear magnetic resonance to lattice quantum chromodynamics, to name a few diverse examples (see Ref.~\cite{fom} and references therein). In many applications, the matrix function $f(M)$  where $M\in\mathbb{C}^{N\times N}$ is the matrix and $f(\cdot)$ is the function, cannot be feasibly computed or stored, or one is not necessarily interested in computing all matrix elements of $f(M)$,  and so specialized algorithms for calculating the action of $f(M)$ on a given vector $|z\rangle$ or the inner product involving two vectors  $\langle y | f(M)| z \rangle$ have been devised.  

In the literature, two main classes of methods have been considered for evaluating $f(M)| z \rangle $.
These are {(i)} {polynomial expansion-based methods} that construct a polynomial 
approximation $p\approx f$ on $M$'s spectral region and then evaluate $p(M)| z \rangle \approx f(M)| z \rangle $, and {(ii)} {Krylov subspace methods} that seek their approximants to $f(M)| z \rangle $ from a Krylov space 
$$\mathcal{K}_q(M, |z\rangle ) := \mathrm{span} \{ | z \rangle, M | z \rangle,\ldots, M^{q-1} | z \rangle\}$$ 
by means of projection. Indeed in many cases, good approximations can be obtained even if $q\ll N$. 
We refer the reader to Ref.~\cite{GKL20} for a recent review of these approaches, and to Ref.~\cite{higham2020catalogue} for a catalogue of software that implements some of the available algorithms. We also note here that there are variants of both method classes mentioned above which are based on rational approximations instead of polynomial ones, leading to {rational Krylov subspace methods}; see Ref.~\cite{Gue13b} for a review. Such methods may converge significantly faster but are only applicable if shifted linear systems with $M$ can be solved efficiently, e.g., via Gaussian elimination or preconditioned iterative solvers, and they also require more parameter tuning than polynomial methods. 

In this study, we propose  another, different, algorithm for calculating individual matrix elements of $f(M)$. We will demonstrate that our algorithm is applicable in cases where the above methods fail because the vectors $M^{j} | z \rangle$ get more and more dense as $j$ increases and can no longer be stored, even if $| z \rangle$ has only a single nonzero entry.    Our approach builds on the recently introduced off-diagonal series expansion~\cite{ODE,ODE2,pmr} which provides a systematic \emph{memory efficient} way of obtaining individual matrix elements by summing a series in which every summand may be interpreted as a walk on a graph.

Certain interpretations of this sort are well known and exploited in the literature, e.g., in the network community; see, e.g., Ref.~\cite[Theorem 2.2.1]{cvetkovic1997eigenspaces} and the review Ref.~\cite{estrada2010network}.
Also, an approach for computing matrix elements of $f(M)$ analytically and in closed form involving summation on graphs (path sums) has been proposed in Ref.~\cite{giscard2013evaluating}. Our approach, on the other hand, is amenable to numerical treatment and, as we shall see, can be implemented efficiently in floating point arithmetic. 
The key idea of our approach that circumvents the sparsity problem encountered with methods that compute the whole vector $f(M)| z \rangle $ is to decompose the matrix $M$ as a sum of (generalized) permutation matrices and to treat each term acting on a sparse vector $| z \rangle$ separately. Because multiplication of a vector by a permutation matrix does not increase sparsity, the scheme can be organized so that it never requires storage of dense vectors. 

The outline of this work is as follows: our method is presented in Sec.~\ref{sec:method}, introducing the off-diagonal expansion approach for matrix functions and applying it for the task of estimating individual entries of $f(M)$. Sec.~\ref{sec:numex} is devoted to numerical experiments demonstrating the viability of our approach, including a demonstration of potential problems encountered with polynomial expansion-based and Krylov methods when the problem size $N$ is extremely large. In Sec.~\ref{sec:denseLin} we consider the application of our method to dense linear systems of equations. A summary is provided in Sec.~\ref{sec:conclusions}.

\section{General method\label{sec:method}}

In this section, we present the off-diagonal series expansion for matrix functions in a given basis. This expansion will serve as the foundation for the proposed algorithm.  

\subsection{The off-diagonal expansion of matrix functions}

Consider a matrix $M\in\mathbb{C}^{N\times N}$ cast in the form
\beq \label{eq:basic}
M=\sum_{j=0}^\ell \widetilde{P}_{j} =\sum_{j=0}^\ell D_j P_j   = D_0 + \sum_{j=1}^\ell D_j P_j \,,
\eeq
where $\{ \widetilde{P}_j\}$ is a set of $\ell+1$ distinct generalized permutation matrices~\cite{gpm}, i.e., matrices with precisely one nonzero element in each row and each column (this condition can be relaxed to allow for zero rows and columns). 
Each matrix $\widetilde{P}_j$ can be written, without loss of generality, as $\widetilde{P}_j=D_j P_j$ where $D_j$ are diagonal matrices and the matrices $P_j$ correspond to permutations with no fixed points (i.e., all diagonal elements are zero) except for the identity matrix $P_0=\mathds{1}$. The diagonal matrix $D_0$ is the diagonal component of $M$. Each term $D_j P_j $ obeys 
$D_j P_j | z \rangle = d_j(z') | z' \rangle$, where $|z\rangle $ can be taken as a canonical unit vector $|z\rangle= [0,\ldots,0,1,0,\ldots,0]^T$,    $d_j(z')$ is a complex-valued coefficient, and $|z'\rangle \neq |z\rangle$ is a basis state of unit norm. The above representation, which we refer to as a `permutation matrix representation' is general and can be applied to any given matrix~\cite{pmr}. 

Now consider the action of $f(M)$ on a basis state $|z\rangle$, assuming that $f(\cdot)$ obeys a Maclaurin series expansion with a region of convergence containing the eigenvalues of $M$:
\bea
f(M)|z\rangle= \sum_{n=0}^{\infty}\frac{f^{(n)}(0)}{n!} M^n | z \rangle
 =\sum_{n=0}^{\infty}\frac{f^{(n)}(0)}{n!}  \Big(D_0+\sum_{j=1}^\ell D_j P_j\Big)^n | z \rangle =  \sum_{n=0}^{\infty}  \sum_{{{\bf i}_n}} \frac{f^{(n)}(0)}{n!}  R_{{\bf i}_n} | z \rangle \,,
\eea
where $f^{(n)}(0)$ is the $n$-th derivative of $f(\cdot)$ at zero  and $\{R_{{\bf i}_n}\}$ denotes the set of all sequences of length $n$ composed of products of basic matrices $D_0$ and $D_j P_j$. 
In the last step we have expressed $M^n$ in terms of all products $R_{{\bf i}_n}$ of length $n$ of basic operators $D_0$ and $D_j P_j$. Here ${\bf i}_n=(i_1,i_2,\ldots,i_n)$ is a set of indices, each of which runs from $0$ to $\ell$, that denotes which of the $\ell+1$ operators in $M$ appear in $R_{{\bf i}_n}$.

We proceed by stripping away all the diagonal Hamiltonian terms from the sequence $ {R}_{{\bf i}_n} | z \rangle$. We do so by evaluating the action of these terms on the relevant basis states, leaving only the off-diagonal operators unevaluated inside the sequence (see Refs.~\cite{ODE,ODE2} for a more detailed derivation). 
The vector $f(M)|z\rangle$ may then be written as
\bea\label{eq:snsq2}
f(M)|z\rangle =
\sum_{q=0}^{\infty} \sum_{\{ {{\bf i}_q}\}}  \left(  \prod_{j=1}^q d^{(i_j)}_{z_j} \right) S_{{\bf{i}}_q} | z \rangle  \left( \sum_{n=q}^{\infty} \frac{f^{(n)}(0)}{n!}  \sum_{\sum k_i=n-q} (E_{z_0})^{k_0} \cdot \ldots \cdot (E_{z_{q}})^{k_{q}} \right)\,,
\eea
where $E_{z_i}=\langle z_i | D_0 | z_i \rangle$, ${\{S_{{\bf{i}}_q}\}}$ denotes the set of all products of length $q$ of `bare' \emph{off-diagonal} operators $P_j$,
and ${{\bf{i}}_q} = (i_1,i_2,\dots,i_q)$ is a set of indices, each of which now runs from $1$ to $\ell$.
Also 
\beq \label{eq:dij}
d^{(i_j)}_{z_j} = \langle z_j | D_{i_j}|z_j\rangle \, ,
\eeq
which can be considered as the `hopping strength' of $P_{i_j}$ with respect to $|z_j\rangle$.

\begin{figure}[htbp]
\includegraphics[width=0.3\textwidth]{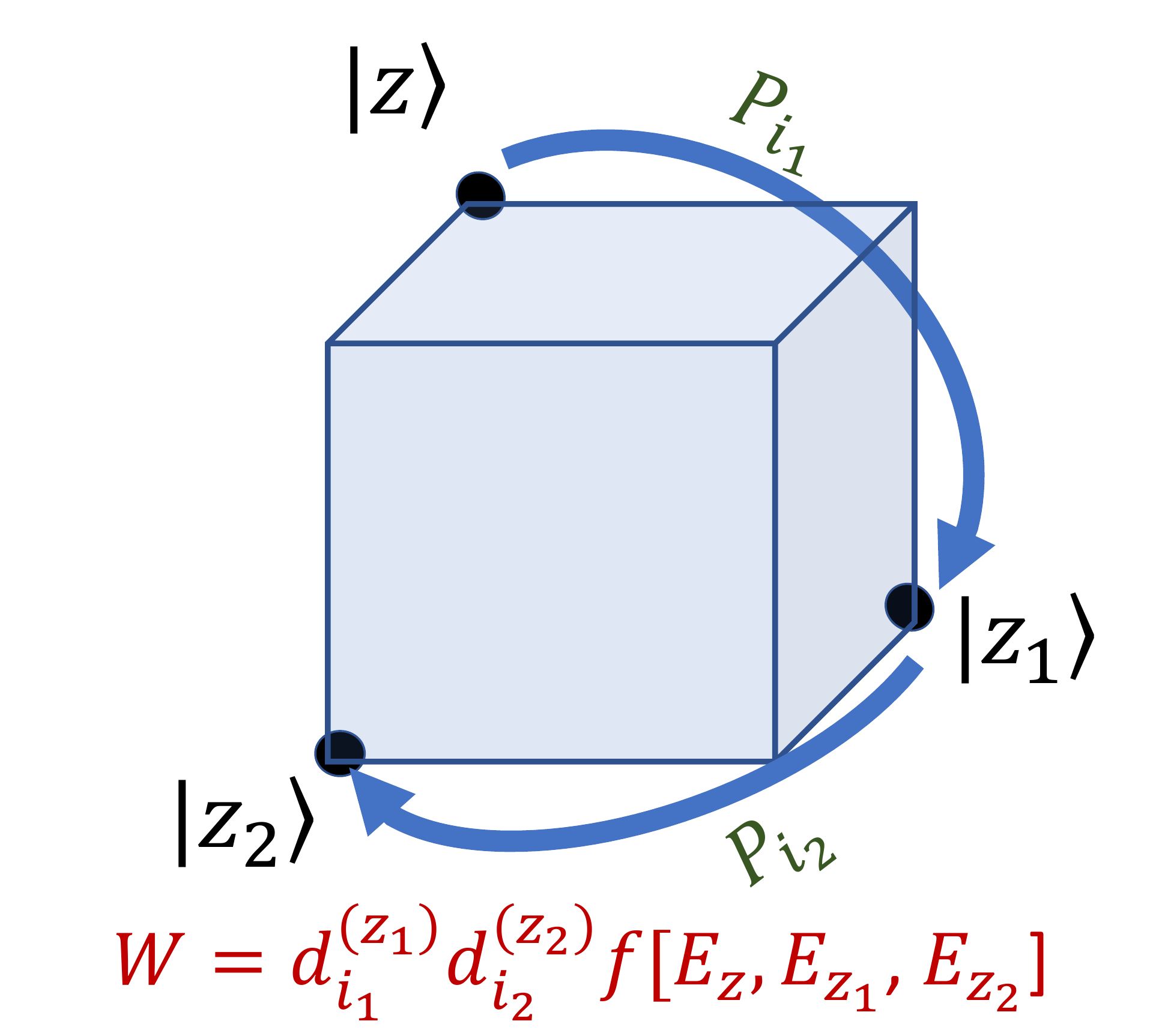}
\small
\cprotect\caption{\label{fig:hyper} A diagrammatic representation of a three-vertex two-edge walk on the graph whose nodes are basis states, starting from the initial basis state $|z\rangle$. Here, $q=2$ and ${\bf i}_q=(i_1,i_2)$. The weight of the walk, which visits the vertices $|z\rangle$, $|z_1\rangle$ and $|z_2\rangle$ is $d_{i_1}^{(z_1)} d_{i_2}^{(z_2)} f[E_z,E_{z_1},E_{z_2}]$.  
} 
\end{figure}

The term in parentheses in Eq.~\eqref{eq:snsq2} sums over the diagonal contribution of all $ R_{{\bf i}_n} | z \rangle$ terms that correspond to the same $ S_{{\bf{i}}_q} | z \rangle$ term. The various $\{|z_i\rangle\}$ states are the states obtained from the action of the ordered $P_j$ operators in the product $S_{{\bf{i}}_q}$ on $|z_0\rangle$, then on $|z_1\rangle$, and so forth. For example, for $S_{{\bf{i}}_q}=P_{i_q} \ldots P_{i_2}P_{i_1}$, we obtain $|z_0\rangle=|z\rangle, P_{i_1}|z_0\rangle=|z_1\rangle, P_{i_2}|z_1\rangle=|z_2\rangle$, etc. The proper indexing of the states $|z_j\rangle$ along the path is \hbox{$|z_{(i_1,i_2,\ldots,i_j)}\rangle$} to indicate that the state at the $j$-th step depends on all $P_{i_1}\ldots P_{i_j}$. We will use the shorthand $|z_j\rangle$. The sequence of basis states $\{|z_i\rangle \}$ may be viewed as a `walk' on the graph whose adjacency matrix is $M$~\cite{ODE,ODE2,signProbODE} (see Fig.~\ref{fig:hyper}).

After a change of variables, $n \to n+q$, we arrive at
\bea
f(M)|z\rangle= \sum_{q=0}^{\infty} \sum_{{\bf i}_q}  S_{{\bf{i}}_q} | z \rangle \left(D_{(z,{\bf{i}}_q)} \sum_{n=0}^{\infty} 
 \frac{f^{(n+q)}(0)}{(n+q)!}  \sum_{\sum k_i=n}  (E_{z_0})^{k_0} \cdots (E_{z_{q}})^{k_{q}} \right),
\eea
where we have also denoted $D_{(z,{\bf{i}}_q)}=\prod_{j=1}^q d^{(i_j)}_{z_j}$.
Noting that the various $\{E_{z_i} \}$ are the diagonal elements associated with the states $|z_i\rangle$ created by the operator product $S_{{\bf{i}}_q}$, the vector $f(M)|z\rangle$ is now given by
\bea \label{eqt:infinitesum}
f(M)|z\rangle= \sum_{q=0}^{\infty} \sum_{ {\bf i}_q} S_{{\bf{i}}_q} | z \rangle D_{(z,{\bf{i}}_q)}
 \left( 
\sum_{\{ k_i\}=(0,\ldots,0)}^{(\infty,\ldots,\infty)} \frac{f^{(q+\sum k_i)}(0)}{(q+\sum k_i)!} \prod _{j=0}^{q} (E_{z_j})^{k_j} 
\right)
 \,.
 \eea
A feature of the above infinite sum is that the term in parentheses can be efficiently calculated as it can be explicitly written as: 
\bea
\sum_{\{ k_i\}} \frac{f^{(q+\sum k_i)}(0)}{(q+\sum k_i)!} \prod _{j=0}^{q} E_{z_j}^{k_j} 
=f[E_{z_0},\ldots,E_{z_q}], 
\eea
where 
\beq 
f[E_{z_0},\ldots,E_{z_q}] \equiv \sum_{j=0}^{q} \frac{f(E_{z_j})}{\prod_{k \neq j}(E_{z_j}-E_{z_k})}
\eeq
are the \textit{divided differences}~\cite{dd:67,deboor:05} of the function $f$
(see Appendix~\ref{app:dd}).
The resultant vector $f(M)|z\rangle$ thus ends up taking the form
\beq 
f(M)|z\rangle = \sum_{q=0}^{\infty} \sum_{\{ {{\bf i}_q}\}} S_{{\bf{i}}_q} | z \rangle D_{(z,{\bf{i}}_q)} f[E_{z_0},\ldots,E_{z_q}] \,. 
\label{eq:fmz}
\eeq
Equation~\eqref{eq:fmz} is the main result of this section. Every summand contributes to a specific basis state, namely $S_{{\bf{i}}_q} | z \rangle$, and can be associated with a walk on the graph defined by the off-diagonal elements of $M$ starting at the basis state $|z\rangle$. The product of permutation matrices $S_{{\bf{i}}_q}=P_{i_q} \cdots P_{i_2} P_{i_1}$ may be viewed as a sequence of hops from $|z\rangle$, to $|z_1\rangle=P_{i_1}|z\rangle$ to $|z_2\rangle=P_{i_2}|z_1\rangle$ and so on. Every $P_{i_j}$ contributes a factor $d_{i_j}^{(z_j)}$ [as per Eq.~(\ref{eq:dij})] and every vertex has an associated diagonal element $E_{z_j}$. The total weight of the walk is 
$D_{(z,{\bf{i}}_q)} f[E_{z_0},\ldots,E_{z_q}]$. This is illustrated in Fig.~\ref{fig:hyper}. 

If the off-diagonal elements are sufficiently small, one can consider the off-diagonal expansion~(\ref{eq:fmz}) as the expression of a perturbation theory, where the specificity of the problem is that the unperturbed part is the diagonal of the matrix $M$, while the perturbation is everything off-diagonal. Here, $q$ is the perturbation order, so Eq.~(\ref{eq:fmz}) contains all orders of perturbation, and the developed expression can be applied to any matrix function, e.g., describing partition function or evolution operator.

\subsection{Evaluating individual matrix elements by summing over walks}
Similar to the evaluation of vectors, individual matrix elements of $f(M)$ can be written as
\beq
\langle z_{\fin} | f(M)|z_{\ini}\rangle  =  \sum_{q=0}^{\infty}  \sum_{{\bf{i}}_q} D_{(z,{{\bf{i}}_q})}  {f [E_{z_0},\ldots,E_{z_q}]} 
\langle z_{\fin} | S_{{\bf{i}}_q} |z_{\ini}\rangle  \,.
\eeq
The term $\langle z_{\fin} | S_{{\bf{i}}_q} |z_{\ini}\rangle$ evaluates to $1$ if $S_{{\bf{i}}_q}$ generates a walk between $|z_{\ini}\rangle$ and $|z_{\fin}\rangle$, and otherwise it is zero. Hence one can rewrite   
\beq\label{eq:matEl}
\langle z_{\fin} | f(M)|z_{\ini}\rangle  =  \sum_{q=0}^{\infty} \sum_{\langle z_{\fin} | S_{{\bf{i}}_q} |z_{\ini}\rangle=1} D_{(z,{{\bf{i}}_q})}  {f [E_{z_0},\ldots,E_{z_q}]}  \,,
\eeq
where the sum is over all walks connecting $|z_{\ini}\rangle$ to $|z_{\fin}\rangle$.

\subsection{Computational considerations}

The off-diagonal representation Eq.~\eqref{eq:matEl} gives us matrix elements $\langle z_{\fin} | f(M)|z_{\ini}\rangle$
as a sum over infinitely many walks. The weight associated with each $q$-edge walk is a product of $q$ matrix elements and one divided difference whose $q+1$ inputs are the diagonal matrix elements associated with the vertices of the walk. We call $q$  the \emph{order} of the expansion. 
Generally, the evaluation of a divided difference requires $O(q^2)$ basic floating point operations~\cite{divdiff2020}. 

While in principle, the order $q$ runs from zero to infinity, we note that the summands obey the divided differences mean value theorem [if $f(\cdot)$ and the diagonal elements are real]:
\beq
{f [E_{z_0},\ldots,E_{z_q}]} =\frac{f^{(q)}(\xi)}{q!}\,,
\eeq
for some $\xi \in [\min E_{z_j}, \max E_{z_j}]$.
Assuming that the Taylor series of $f(\xi)$ converges sufficiently fast, 
it is  allowed in practice to cut off the series Eq.~(\ref{eq:matEl}) at a properly chosen 
maximal order $Q$ without incurring any significant error.
The value of $Q$ determines the time and memory cost of the algorithm.
This value depends on the choice of function $f(\cdot)$, so it is analyzed
separately for each case in the next sections.

Denoting by $\Gamma$ a bound on the modulus of the off-diagonal elements, we find that the contribution of a walk is
roughly $\Gamma^q f^{(q)}(\xi)/q!$ meaning that an appropriate choice for $Q$ is proportional to $\Gamma$. 
This further implies that the cost of the algorithm 
grows rapidly with the increase of the norm of the off-diagonal component of the matrix.
Hence a practical prerequisite for applying the method is that the off-diagonal matrix elements are sufficiently small. 
 
Another requirement is that the matrix $M$ should be sufficiently sparse 
so that the number of different walks is not too large for the feasibility of considering
all summands in Eq.~(\ref{eq:matEl}).
An exception is the case when there are only few distinct diagonal elements, 
and a contribution $D_{(z,{{\bf{i}}_q})} {f[E_{z_0},\ldots,E_{z_q}]}$ in Eq.~(\ref{eq:matEl})
can only take on a few different values. In this case the number of walks
resulting in each possible value of the contribution should be computed
rather than considering each walk separately.
 
In the next sections, we present the algorithm in detail for several important functions and matrix classes and investigate the method numerically.
We will show that we can calculate to machine accuracy the matrix elements of functions
of extremely large matrices (up to size $2^{64} \times 2^{64}$ on a single workstation) for which existing methods simply cannot provide results. 

\section{Calculating elements of the matrix exponential}\label{sec:numex}

In this section we focus on the  exponential function $f(x) = \exp(x)$ and its variants $f(x)=\exp(-\beta x)$ and $f(x)=\exp(-itx)$ where $\beta$ and $t$ are real-valued. 
We successfully compute off-diagonal elements of  matrices $M$ of sizes  $2^n \times 2^n$, where $n=L^2$ and $L$ is as large as $8$, on a single workstation. A crucial assumption is that  the off-diagonal  elements in $M$ are sufficiently small. We also review in Sec.~\ref{sub:poly}  
alternative polynomial approximation methods which are commonly 
used for the same task and discuss why they  struggle to solve such problems.

Our focus is a class of matrices that appear widely in physics applications, specifically condensed matter physics, namely the transverse-field Ising Hamiltonians~\cite{tfim1,tfim2}. These have the general form 
\begin{equation}
M = \sum_{\langle i,j\rangle} J_{ij} Z_i Z_j - \Gamma \sum_j X_j,
\label{hamiltonian}
\end{equation}
where the two-dimensional lattice of size $L\times L$ containing $n=L^2$ spins is considered.
We use $\langle i,j\rangle$ 
in Eq.~(\ref{hamiltonian}) to denote that $i$ and $j$ are neighboring spins 
on a 2D lattice with periodic boundary conditions. Here, $J_{ij}$ are real-valued parameters and
$X_j$ and $Z_j$ are the Pauli-$x$ and Pauli-$z$ matrices, respectively, acting on the $j$-th spin. 
Hence, $M$ is a sparse symmetric matrix of size $2^{n} \times 2^{n}$; each row or column
of the matrix contains $n$ off-diagonal nonzero elements, which are equal to $-\Gamma$.
Expressed differently, if we label basis vectors using an $n$-bit binary representation 
then the matrix $Z_i Z_j$ is a diagonal matrix with 1 in the $(k,k)$-th 
entry if the $i$-th and $j$-th bits of the index $k$ are the same. 
Otherwise, the $(k,k)$-th entry is $-1$. The off-diagonal matrix $X_j$ 
is zero everywhere except for the $(k,k')$-th entries wherever the bit 
representation of $k$ and $k'$ differ (only) in the $j$-th bit, in which case the entry is 1. 

We start from considering
a simplified but nontrivial variant of the above model where the Ising part of the transverse-field Hamiltonian is taken with $J_{ij}=1/4$ and modulo 2:
\begin{equation}
M = \frac14 \sum_{\langle i,j\rangle} Z_i Z_j \Mod{2} - \Gamma \sum_j X_j.
\label{hamiltonianmod2}
\end{equation}
In this case, the matrix diagonal contains only zeros and ones and hence
the computation of the  divided differences of the diagonal entries 
is limited to a relatively small number of different values, 
significantly speeding up the computation~\footnote{If $L$ is odd,
the first summand in~(\ref{hamiltonianmod2}) should be understood as
$\left\lfloor \frac14 \left| \sum_{\langle i,j\rangle} Z_i Z_j \right| \right\rfloor\Mod{2}$,
where $\lfloor x\rfloor$ denotes the largest integer that is less than or equal to $x$.}.  As a consequence, the matrix elements 
of $\e^{-\beta M}$, the trace of which is called the partition function of $M$ for inverse temperature $\beta$, can be computed for larger values of $\Gamma$
as compared to the full model Eq.~\eqref{hamiltonian}.

We have made our program codes available on GitHub~\cite{MatrixFunctionsCode}.

\subsection{Transverse-field Ising model (TFIM) modulo two}
\label{sec:isingmod2}

In this section, we consider the matrix $M$ defined in Eq.~(\ref{hamiltonianmod2}).
The resulting algorithm can accurately obtain individual matrix elements
of $f(M) = \exp(-\beta M)$, where $M$ is defined in Eq.~\eqref{hamiltonianmod2} and the value 
of $\beta\Gamma$ is not large.
Numerically, we have obtained several matrix elements of $\exp(-\beta M)$,
where we used the parameters $L=8$, $\beta=1$, $\Gamma=0.05$. 

\subsubsection{Calculating the number of walks}
\label{sec:numberofwalks}

Here, we calculate $W(q,m)$, the number of walks $S_{\mathbf{i}_q}$ of length $q$ between $z_\ini$ and $z_\fin$ for the above model, where $m$ is the number of different bits between $\ini$ and $\fin$. To be specific, we assume here and below that the particular bits that differ are the $m$ least significant bits. This requirement does not result in a loss of generality of the relations below, and it is easy to get rid of this assumption in the program code. By definition,
\beq
\label{eq:Nqsum}
W(q,m) = \sum_{\substack{k_1+\ldots+k_n = q \\ k_1,\cdots,k_m \textrm{odd} \\ k_{m+1},\ldots,k_n \textrm{even}}}
{\binom{q}{k_1 k_2 \ldots k_n}},
\eeq
where $k_j$ is the number of times that matrix $X_j$ appears in the sequence.
We note that $\binom{q}{k_1 k_2 \cdots k_n}$ is the coefficient of $x_1^{k_1}\cdots x_n^{k_n}$
in the expansion of $(x_1+\cdots+x_n)^q$. The sum of all these coefficients is
obtained by substituting $x_1=\cdots=x_n=1$. 
We note that only even powers $k_1$ are preserved in the expansion of $\frac12 [(x_1+\cdots+x_n)^q + (-x_1+\cdots+x_n)^q]$; 
only odd powers $k_1$ are preserved in the expansion of $\frac12 [(x_1+\cdots+x_n)^q - (-x_1+\cdots+x_n)^q]$.
Therefore,
\beq
W(q,m) = \frac1{2^n}\sum_{v_i = -1,1} v_1v_2\cdots v_m \left(v_1+v_2+\cdots+v_n\right)^q,
\eeq
which can be further simplified to
\beq
W(q,m) = \frac1{2^n}\sum_{k=0}^{n-m}\sum_{r=0}^m {\binom{m}{r}}{\binom{n-m}{k}} (-1)^r (n-2k-2r)^q.
\eeq
For the case $m=0$, we have
\beq
W(q,0) = \frac1{2^n}\sum_{k=0}^n {\binom{n}{k}} (n-2k)^q.
\label{eq:Nqdiag}
\eeq

\subsubsection{Generation of all walks}
\label{sec:walkGeneration}

The generation of all $W(q,m)$ walks for a given $q$ consists of the following two tasks.

\begin{enumerate}

\item
Generation of all possible sets of values $k_1,\ldots,k_n \geq 0$ such that
$k_1+\cdots+k_n=q$, $k_1,\dots,k_m$ are odd, and $k_{m+1},\dots,k_n$ are even.

Given that $k_1+\cdots+k_n=q$ is equivalent to
\beq
\frac{k_1+1}{2}+\cdots+\frac{k_m+1}{2}+\left(\frac{k_{m+1}}{2}+1\right)+\cdots+\left(\frac{k_n}{2}+1\right) =\frac{q-m}{2}+n,
\eeq
one needs to find $n$ positive integers such that their sum is equal to $(q-m)/2+n$.
This is equivalent to placing $n-1$ walls between $(q-m)/2+n$ items.
Hence, the task is equivalent to generating all subsets of size $n-1$ of a set $\{1,2,\ldots,w\}$, where
$w=(q-m)/2+n-1$. There are $\binom{w}{n-1}$ such subsets.
In order to generate all such subsets, we start from the subset $\{1,2,\dots,n-1\}$ and then successively
increase its elements from right to left, preserving the increasing order of the elements.

\item
Generation of all walks such that the number of times that matrix $X_j$
appears in the sequence is equal to $k_j$ for $j=1,2,\ldots,n$ and for given values of $k_1,\ldots,k_n$.
The pseudocode for the recursive implementation of this routine, which employs the calculation of divided differences 
by addition and removal of inputs~\cite{divdiff2020}, is shown in listing~\ref{routine}.

\end{enumerate}

\begin{algorithm}[H]
\caption{Generation of all walks such that the number of flips of spin number $j$ is $k_j$
for $j=1,2,\ldots,n$.}
\label{routine}
\begin{algorithmic}[1]
\Procedure{WalkGeneration}{$k_1,k_2,\dots,k_n$}
	\State {\bf Data stored in global variables:}
	\State \algspace current length $l$,
	\State \algspace sequence of energies $E_{z_j}=\langle z_j | M | z_j \rangle$ for $j=0,1,2,\dots,l$,
	\State \algspace divided differences $\exp[-\beta E_{z_0},\ldots,-\beta E_{z_j}]$ for $j=0,1,2,\dots,l$,
	\State \algspace the configuration vector which stores the states of $n=L^2$ spins and determines the current basis state.
	\State {\bf Initialization before executing the routine:}
	\State \algspace the initial length $l=0$;
	\State \algspace the configuration vector initially corresponds to the basis state $|z_\ini\rangle$;
	\State \algspace the sequence of energies contains $E_{z_\ini}$;
	\State \algspace the sequence of divided differences contains $\exp[-\beta E_{z_\ini}]$.
	\If{$ l = q$}
		\State process the current walk, i.e., add the current divided difference to the sum.
	\Else
		\ForAll{$i$ such that $k_i>0$}
			\State perform the spin flip of $i$-th spin;
			\State calculate the new energy corresponding to the new configuration vector and add it to the list of energies;
			\State calculate the new divided difference by calling the addition routine~\cite{divdiff2020}; 
			\State $l := l + 1$;
			\State $k_i := k_i - 1$;
			\State \Call{WalkGeneration}{$k_1,k_2,\dots,k_n$}
			\State remove the last energy from the sequence of energies;
			\State return to the previous divided difference by calling the removal routine~\cite{divdiff2020};
			\State perform the spin flip of $i$-th spin again;
			\State $l := l - 1$;
			\State $k_i := k_i + 1$. 
		\EndFor
	\EndIf
\EndProcedure
\end{algorithmic}
\end{algorithm}

\subsubsection{Calculating matrix elements of \texorpdfstring{$f(M)$}{}}
\label{sec:matexpalgmod2}

It follows from Eq.~(\ref{eq:matEl}) for the choice $f(x) = \exp(-\beta x)$ that
\beq
\langle z_{\fin} | f(M)|z_{\ini}\rangle  =  \sum_{q=0}^\infty 
\sum_{\langle z_{\fin} | S_{{\bf{i}}_q} |z_{\ini}\rangle=1} (\beta\Gamma)^q  \exp[-\beta E_{z_0},\ldots,-\beta E_{z_q}]  \,.
\label{eq:matexp}
\eeq
In order to compute the sum Eq.~(\ref{eq:matexp}), we follow routines derived in Refs.~\cite{divdiff2020,Zivcovich2019} and 
 initialize the array $d_0,d_1,\dots,d_{q+1}$ with the divided differences 
\beq
d_i = \exp[\underbrace{0,0,\ldots,0}_\text{$q+1-i$ times},\underbrace{-\beta,-\beta,\ldots,-\beta}_\text{$i$ times}].
\eeq
The next step is the generation of all walks of length $q$ as 
described in Sec.~\ref{sec:walkGeneration},
where the calculation of corresponding divided differences should be omitted.
The array $u_0,u_1,\dots,u_{q+1}$ is initialized with zeros for each $q$.
For each walk, we increment $u_i$ by one, where $i$ is the number of nonzero elements among $E_{z_0},\ldots,E_{z_q}$.
Finally, we obtain
\beq
\label{eq:algsummod2}
\langle z_{\fin} | f(M)|z_{\ini}\rangle  = \sum_{q=0}^\infty (\beta\Gamma)^q \sum_{i=0}^{q+1} u_i d_i.
\eeq
The values of $q$ are incremented until the corresponding contribution in Eq.~(\ref{eq:algsummod2}) falls below a given truncation tolerance.

\subsubsection{Computational complexity and memory requirements}
\label{sec:complexityisingmod2}

Since the matrix diagonal contains only zeros and ones, we have
$e^{-\beta}/q! \leq \exp[-\beta E_{z_0}, \dots, -\beta E_{z_q}] \leq 1/q!$~\cite{Farwig1985},
so all divided differences are of the order of $1/q!$ for $\beta=1$.
Therefore, it follows from Eq.~(\ref{eq:matexp}) that the condition
\beq
\frac{W(q,m) (\beta\Gamma)^q}{q!} \lesssim \varepsilon
\label{eq:complexityconditionmod2}
\eeq
needs to be satisfied for all $q>Q$ in order to compute a matrix element, 
where $\varepsilon$ is the error tolerance and $Q$ is the maximal order.
If $q$ is sufficiently large, then it follows from Eq.~(\ref{eq:Nqdiag}) that $W(q,0) \sim n^q$. 
In this case, the maximum  of the left-hand side in Eq.~(\ref{eq:complexityconditionmod2})
is approximately at $q \approx n\Gamma$.
Table~\ref{tab:isingmod2degree} shows that diagonal matrix elements
can be computed on a single workstation
for $L=8$, $\beta=1$, $\Gamma \leq 0.05$ and $\varepsilon = 10^{-8}$.
We have also confirmed this numerically. 
The algorithm requires only $O(n+Q)$ bytes of memory, where $Q$ is the maximal order. 

\begin{table}
\begin{center}
\begin{tabular}{l|l|l|l}
$\Gamma$ & $Q$ & $N_Q$ & time (sec.)\\
\hline
\hline
$1$        &    $142$   &    $6.4\cdot 10^{237}$ &  $3.5\cdot 10^{231}$ \\
$0.1$      &    $14$    &    $4.8\cdot 10^{17}$  &  $2.6\cdot 10^{11}$  \\
$0.05$     &    $10$    &    $9.1\cdot 10^{11}$  &  $5.1\cdot 10^5$     \\
$0.01$     &    $4$     &    $1.2\cdot 10^{4}$   &  $5.0\cdot 10^{-3}$  \\
\end{tabular}
\caption{
Degree $Q$ and the corresponding number of walks $N_Q$
required to approximate a diagonal matrix element with Eq.~(\ref{eq:matexp}),
where the error tolerance is $\varepsilon=10^{-8}$ and $n=64$.
The values are obtained using Eq.~(\ref{eq:complexityconditionmod2}).
\label{tab:isingmod2degree}}
\end{center}
\end{table}

\subsection{The full transverse-field Ising model \label{sec:ising}}

In this section, we consider the matrix $M$ defined in Eq.~\eqref{hamiltonian}, i.e., the full transverse-field Ising model, 
rather than Eq.~\eqref{hamiltonianmod2} and
modify accordingly the algorithm described above in Sec.~\ref{sec:isingmod2}.
The resulting algorithm can accurately obtain individual matrix elements
of $f(M) = \exp(-\beta M)$, where $M$ is defined in Eq.~\eqref{hamiltonian} and the value 
of $\beta\Gamma$ is not large.

We note that the both models involve the same walks. Therefore, the considerations discussed in
Secs.~\ref{sec:numberofwalks} and~\ref{sec:walkGeneration}
fully apply to the present case as well.
However, Secs.~\ref{sec:matexpalgmod2} and~\ref{sec:complexityisingmod2}
should be modified as follows.
(i) The matrix elements are obtained with Eq.~(\ref{eq:matexp}) rather than with Eq.~(\ref{eq:algsummod2}).
Here, the generation of walks described above in Sec.~\ref{sec:walkGeneration} results in
the calculation of corresponding summands of Eq.~(\ref{eq:matexp}) on the fly.
The values of $q$ are incremented until the corresponding 
contribution falls below a given truncation tolerance.
(ii) The diagonal elements $E_{z_0},\dots,E_{z_q}$ can vary strongly,
so the values of divided differences in Eq.~(\ref{eq:matexp}) can vary by many orders of magnitude.
This complicates the theoretical assessment of the maximal order $Q$ for this problem.

\begin{figure}[tb]
\includegraphics[width=0.5\textwidth]{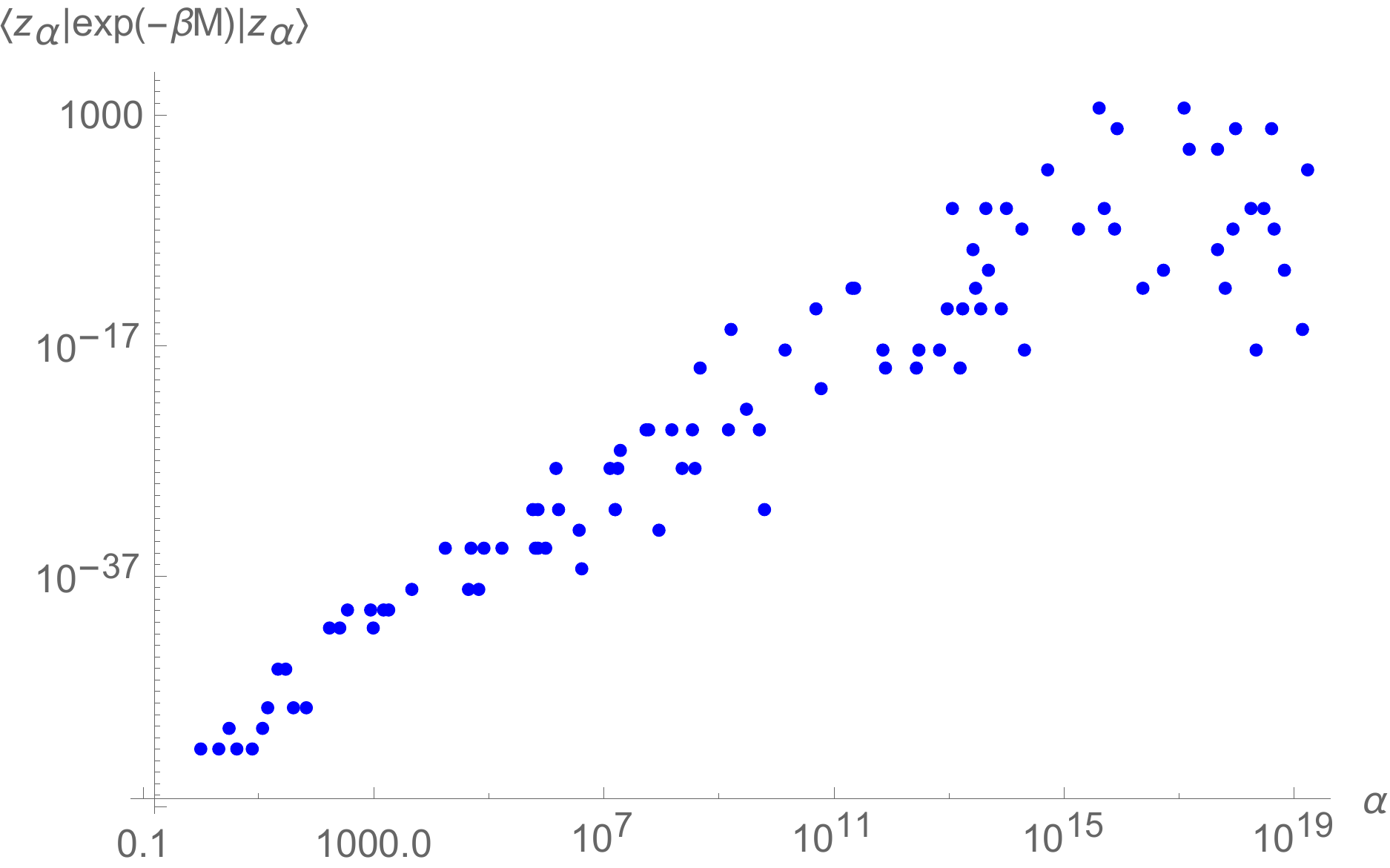}
\caption{\label{fig:tfimdiag} Diagonal matrix elements 
$\langle z_\ini | \exp(-\beta M) | z_\ini \rangle$, where $M$ is the transverse-field Ising model~\eqref{hamiltonian} with
$L=8$, $\beta=1$, $\Gamma=0.01$, $J_{ij}=1$, and $98$ indices $\ini$ were chosen randomly
such that $\log_2 \ini$ was uniformly distributed on the interval $[0,64)$.} 
\end{figure}

\begin{table}
\begin{center}
\begin{tabular}{l|l|l|l|l|l}
\multicolumn{2}{c|}{ }  & \multicolumn{2}{c|}{\ TFIM(mod 2), $\Gamma=0.05$\ \ } &  \multicolumn{2}{c}{TFIM, $\Gamma=0.001$} \\
\hline
$L$ & $N$ & $q$ & $\text{err}(L)$ & $q$ & $\text{err}(L)$   \\
\hline
1 & 2 & 7 &\  $9.933\cdot 10^{-10}$ & 11 & $4.562\cdot 10^{-9}$ \\
2 & 16 & 7 & $6.878\cdot 10^{-9}$ & 23 & $1.775\cdot 10^{-9}$ \\
3 & 512 & 8 & $4.218\cdot 10^{-9}$ & 39 & $2.556\cdot 10^{-9}$ \\
4 & $6.6\cdot 10^{4}$ & 9 & $4.693\cdot 10^{-9}$ & 60 & $6.321\cdot 10^{-9}$ \\
5 & $3.4\cdot 10^{7}$ & 10 & $7.495\cdot 10^{-9}$ & 88 & $2.711\cdot 10^{-9}$ \\
6 & $6.9\cdot 10^{10}$ & 12 & $1.326\cdot 10^{-9}$ & 121 & $4.042\cdot 10^{-9}$ \\
7 & $5.6\cdot 10^{14}$ & 13 & $3.745\cdot 10^{-9}$ & 160 & $5.722\cdot 10^{-9}$  \\
8\ \ \ &  $1.8\cdot 10^{19}$\ \ \ &  15\ \ \ \  & $1.349\cdot 10^{-9}$ &  205\ \ \  &$8.080\cdot 10^{-9}$\ \ \ \ \  \\ 
\end{tabular}
\caption{Polynomial degree $q$ required to uniformly approximate $f(x) = \exp(-x)$ on the intervals $[-n\Gamma, 1+n\Gamma]$ and $[-2n-n\Gamma, 2n+n\Gamma]$ for TFIM(mod 2) and TFIM, respectively. The achieved accuracy $\text{err}(L)$ is below $10^{-8}$ in all cases. Here,  $n=L^2$ and the matrix size $N=2^n$ is also listed. The values for TFIM, $L\geq 3$ are estimated via truncated Chebyshev expansion while the others are computed using the Remez algorithm.\label{tab:degree}}
\end{center}
\end{table}

Our numerical tests show that the computation is feasible 
for sufficiently small values of $\Gamma$.
In particular, we have calculated matrix elements of $\exp(-\beta M)$ on a single workstation 
for $L=8$, $\beta=1$, $\Gamma=0.01$, $J_{ij}=1$, where the error tolerance was
$\varepsilon = 10^{-8}$.
Similar to the previous case, the algorithm requires only $O(n+Q)$ bytes of memory, where $Q$ is the maximal order. 
As a calculation example, Fig.~\ref{fig:tfimdiag} shows several diagonal matrix elements 
$\langle z_\ini | \exp(-\beta M) | z_\ini \rangle$,
where $\ini$ were chosen randomly such that $\log_2 \ini$ was uniformly distributed on the interval $[0,64)$.

\subsection{Comparison to polynomial approximation methods}\label{sub:poly}

In order to put our computational results in context, we first discuss methods based on  polynomial approximation, $p(M)|z\rangle \approx f(M) |z\rangle$, so-called explicit expansion-based methods and Krylov subspace methods. 
These are among the most commonly used approaches for approximating the action of a matrix function $f(M)$ on a vector $|z\rangle$ without requiring the computation of the (generally dense) matrix $f(M)$; see, e.g., \cite{GKL20} for a recent survey of such methods. The viability of such approaches hinges on the ability to compute and store the vectors $M^j |z\rangle$ for $j=0,1,\ldots$, which is implemented by repeated matrix-vector multiplication $M(M\cdots (M|z\rangle))$. Note that a single matrix element  $(f(M))_{ij}$ can be computed by setting $|z\rangle=|e_j\rangle$, the $j$th canonical unit vector, and then extracting the $i$th entry of $f(M)|z\rangle$. 

Recall that each row and column of the matrices in Eq.~\eqref{hamiltonian} (referred to as TFIM) and the simplified model Eq.~\eqref{hamiltonianmod2} (TFIM mod 2) contain $n=L^2$ nonzeros off the diagonal. Starting with a canonical unit vector $|z\rangle$, the number of nonzeros in the vector $M^j |z\rangle$ is therefore bounded by $\max\{ (n+1)^j, 2^n \}$. This may easily exceed the limits of state-of-the-art numerical computing environments. For example, in MATLAB 2019A, it is not even possible to allocate a sparse zero vector of size $2^n \times 1$ if $L>6$, as in this case $2^n > 2^{48}-1$, the maximum number of elements allowed in an array. 

For the TFIM model, Eq.~\eqref{hamiltonian}, $M$ is symmetric with diagonal elements in the interval $[-2n,2n]$, and since there are $n$ off-diagonal elements $-\Gamma$, by Gershgorin's circle theorem we know that the eigenvalues are contained in the interval $[-2n-n\Gamma, 2n+n\Gamma]$. 
For the simplified model TFIM(mod 2), given by Eq.~\eqref{hamiltonianmod2}, $M$ is symmetric with diagonal elements in the interval $[0,1]$, and since there are $n$ off-diagonal elements $-\Gamma$, we know that the eigenvalues of $M$ are contained in the interval $[-n\Gamma, 1+n\Gamma]$. 

For any (polynomial) approximation $p(M)|z\rangle \approx  f(M) |z\rangle$ with the eigenvalues of $M$ contained in the interval $[\mu,\nu]$ we  have the following bound on the element-wise error:
\begin{eqnarray*}
    \| f( M)|z\rangle  - p(M)|z\rangle \|_\infty 
       &\leq& \| f(M)|z\rangle  - p(M)|z\rangle \|_2 \\
       &\leq& \| |z\rangle \|_2 \cdot  \| f(D)  - p(D) \|_2 \\
       &\leq& \max_{x \in [\mu, \nu]} | f(x) - p(x) |,
\end{eqnarray*}
where $D$ is a diagonal matrix containing the eigenvalues of $M$. 
The right-hand side allows us to bound the degree of $p$ required to achieve a chosen element-wise accuracy. One approach, taken in Ref.~\cite{druskin1989two}, is to use for $p$ a degree $q$ Chebyshev expansion of $f(\cdot)$ and to bound the error $|f-p|$ by bounding the Chebyshev coefficients of order $q+1,q+2,\ldots$. For our practical purposes, we numerically calculate the smallest degree $q$ such that 
$$
\text{err}(L) := \min_{p\in\mathcal P_q} \max_{x \in [\mu, \nu]} | f(x) - p(x) | \leq 10^{-8},
$$
a value suitable for many practical purposes, where  the minimum is taken over all polynomials of degree at most $q$, and $f(x) = \exp(-x)$ (i.e., $\beta = 1$). The solution of this polynomial best approximation problem is performed using the Remez algorithm as implemented in the Chebfun package~\cite{driscoll2014chebfun}, or for large intervals $[\mu,\nu]$ where the growth of $f(x)$ prevents a stable numerical computation of a uniform best approximant in double precision it is approximately solved using a truncated Chebyshev expansion. The results are listed in Table~\ref{tab:degree}. Note how $q$ grows relatively slowly as $L$ is increased, thanks to the superlinear convergence of the polynomial best approximation to the exponential function.

Let $A_j$ be the number of nonzero entries in $M^j |z_k\rangle$, 
where the matrix $M$ corresponds to TFIM or TFIM(mod 2), and $|z_k\rangle$ 
is the $k$-th canonical unit vector, $k\in \{0,1,\dots,2^n-1\}$.
Then the following statements hold:
(i) $A_j \geq \binom{n}{j}$ for all $j$; (ii) $A_j \geq 2^{n-1}$ for $j\geq n-1$.
These statements follow from the fact that
$(M^j |z_k\rangle)_i = \sum_{m_1,\dots,m_{j-1}} M_{i,m_1}M_{m_1,m_2}\dots M_{m_{j-2},m_{j-1}}M_{m_{j-1},k}$
is nonzero whenever the bit representation of $i$ can be obtained from the bit representation of $k$
by a change of $j$ bits (any of the bits can change repeatedly).

Hence, the number of nonzero entries in $M^j |z_k\rangle$ grows rapidly with $j$,
and it becomes intractable to store these vectors when the degree $j$ approaches the values of $q$ listed in Table~\ref{tab:degree}. In addition, the cost of one matrix-vector multiplication 
$M^j |z\rangle \rightarrow M^{j+1} |z\rangle$ grows by the same rate (as it scales linearly with the number of nonzeros in $M^j |z\rangle$). All polynomial expansion-based  and Krylov subspace based methods require the storage of at least one such vector, and most commonly, a small number thereof. For example, the two-pass Lanczos method requires the storage of four vectors of the  original problem size $N$, of which  three vectors are part of the Krylov basis $|v_j \rangle$ (and are hence increasingly dense as the iteration number $j$ increases) and a fourth vector is to store the approximation to $f(M)|z\rangle$. 
(If only a single element of $f(M) | z \rangle$ needs to be computed, the memory requirement of the two-pass Lanczos method reduces to essentially three vectors.)
As a consequence, if only $L$ and $q$ are large enough, polynomial expansion-based and Krylov subspace methods will suffer from increased computational cost with each iteration and soon exhaust all available memory even for just storing a single vector. 

\subsection{Accuracy, wall-clock time and memory consumption of the proposed algorithm}

\begin{figure}[t]
\includegraphics[width=0.8\textwidth]{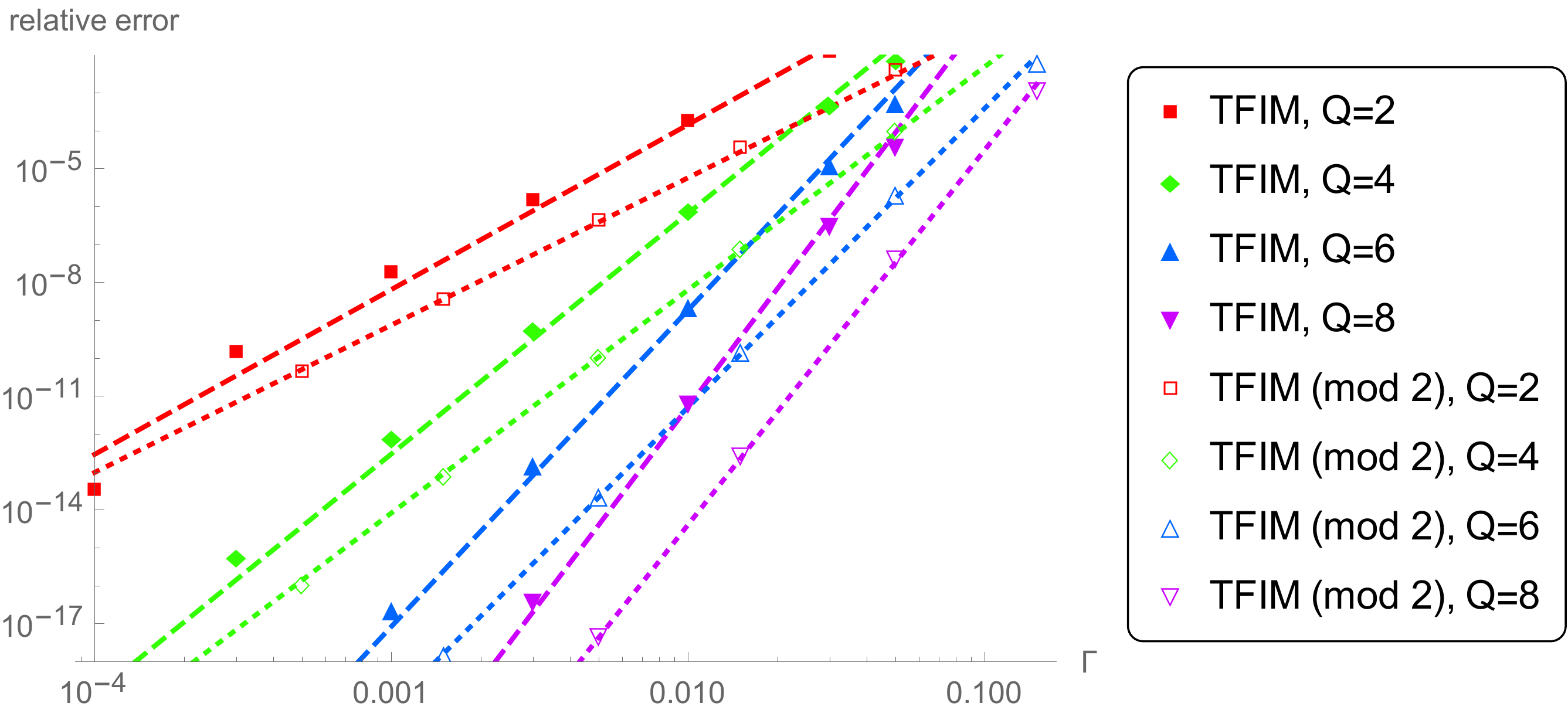}
\caption{\label{fig:error} Estimated relative error of the computed value of $\langle z_\ini |\exp(-\beta M) | z_\ini\rangle$ 
as a function of the off-diagonal strength $\Gamma$ and maximal expansion order $Q$.
Here, $\ini=16210525687446977967$.} 
\end{figure}

\begin{figure}[t]
\includegraphics[width=0.8\textwidth]{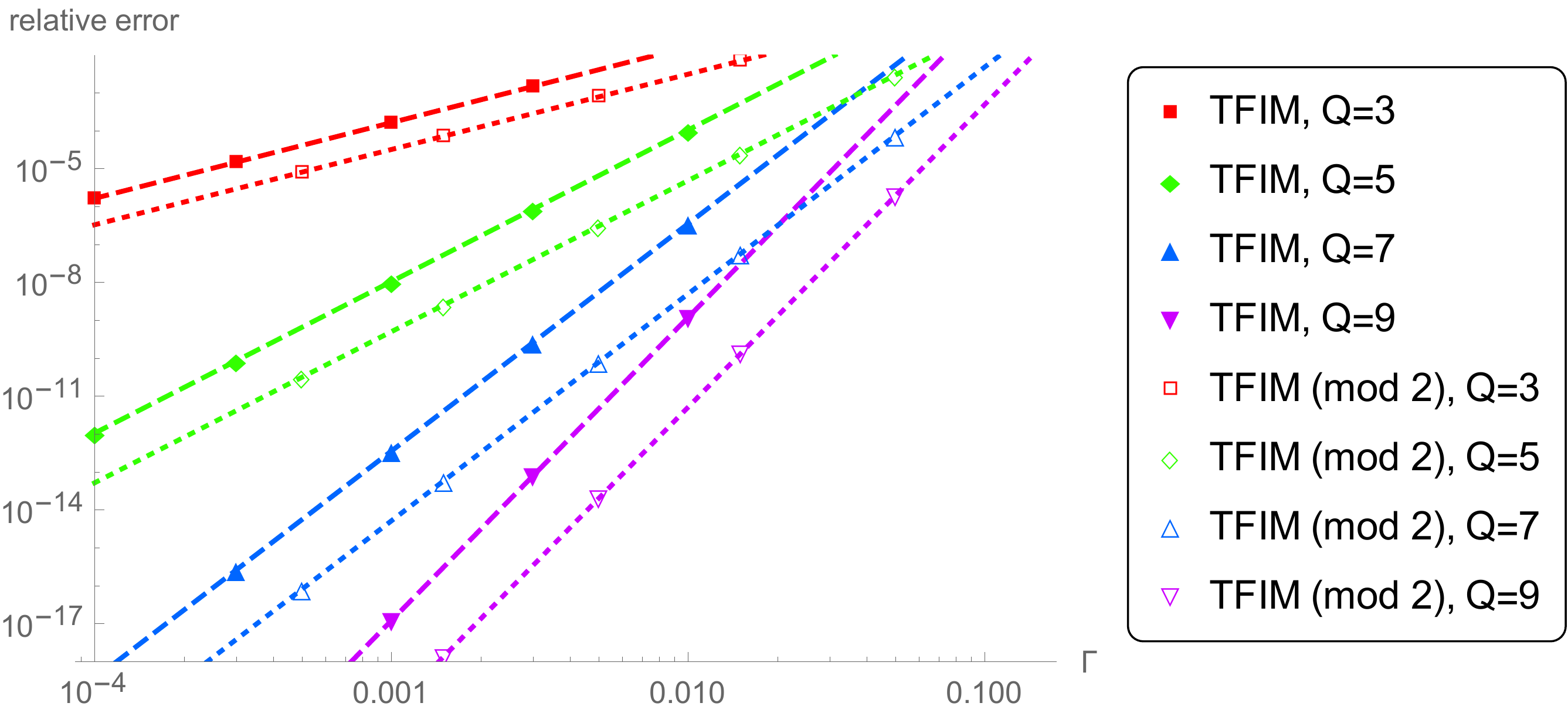}
\caption{\label{fig:error:nondiag} Estimated relative error of the computed 
nondiagonal matrix element $\langle z_\fin |\exp(-\beta M) | z_\ini\rangle$ 
as a function of the off-diagonal strength $\Gamma$ and maximal expansion order $Q$.
Here, $\ini=16210525687446977967$, $\fin=16209397588516748719$, so the number of different bits
between $\ini$ and $\fin$ is $m=3$.} 
\end{figure}

\begin{figure}[tbh]
\includegraphics[width=0.6\textwidth]{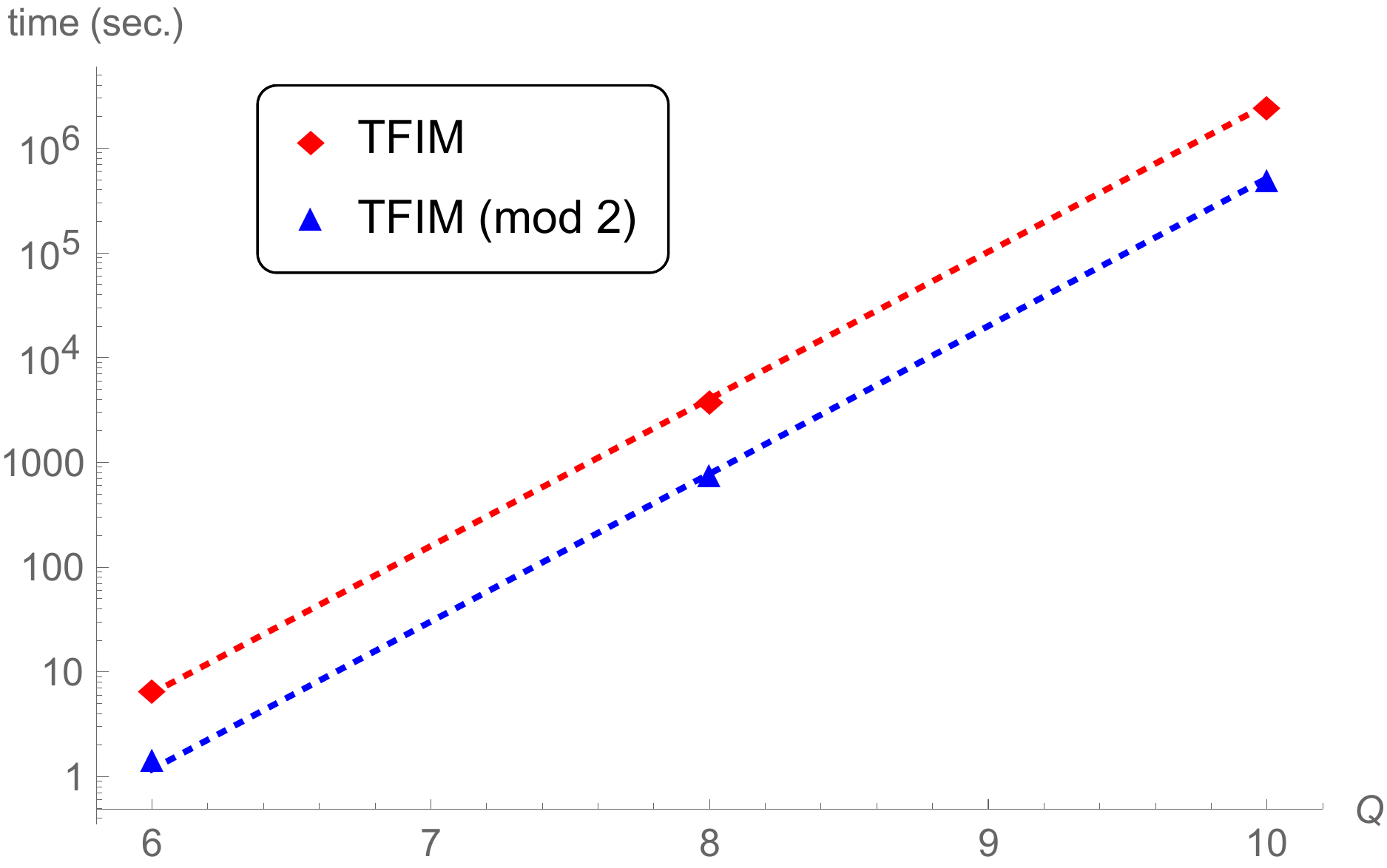}
\caption{\label{fig:time} Wall-clock time of computing $\langle z_\ini |\exp(-\beta M) | z_\ini\rangle$
as a function of $Q$. The dotted lines correspond to
$2.33\cdot 10^{-8}\cdot 25.383^Q$ and $4.03\cdot 10^{-9}\cdot 25.737^Q$ for TFIM and TFIM $\Mod{2}$, correspondingly.
For $Q\gtrsim 100$, the time is proportional to $n^Q$, where $n=64$ for the considered case.}
\end{figure}

\begin{figure}[htp]
\includegraphics[width=0.6\textwidth]{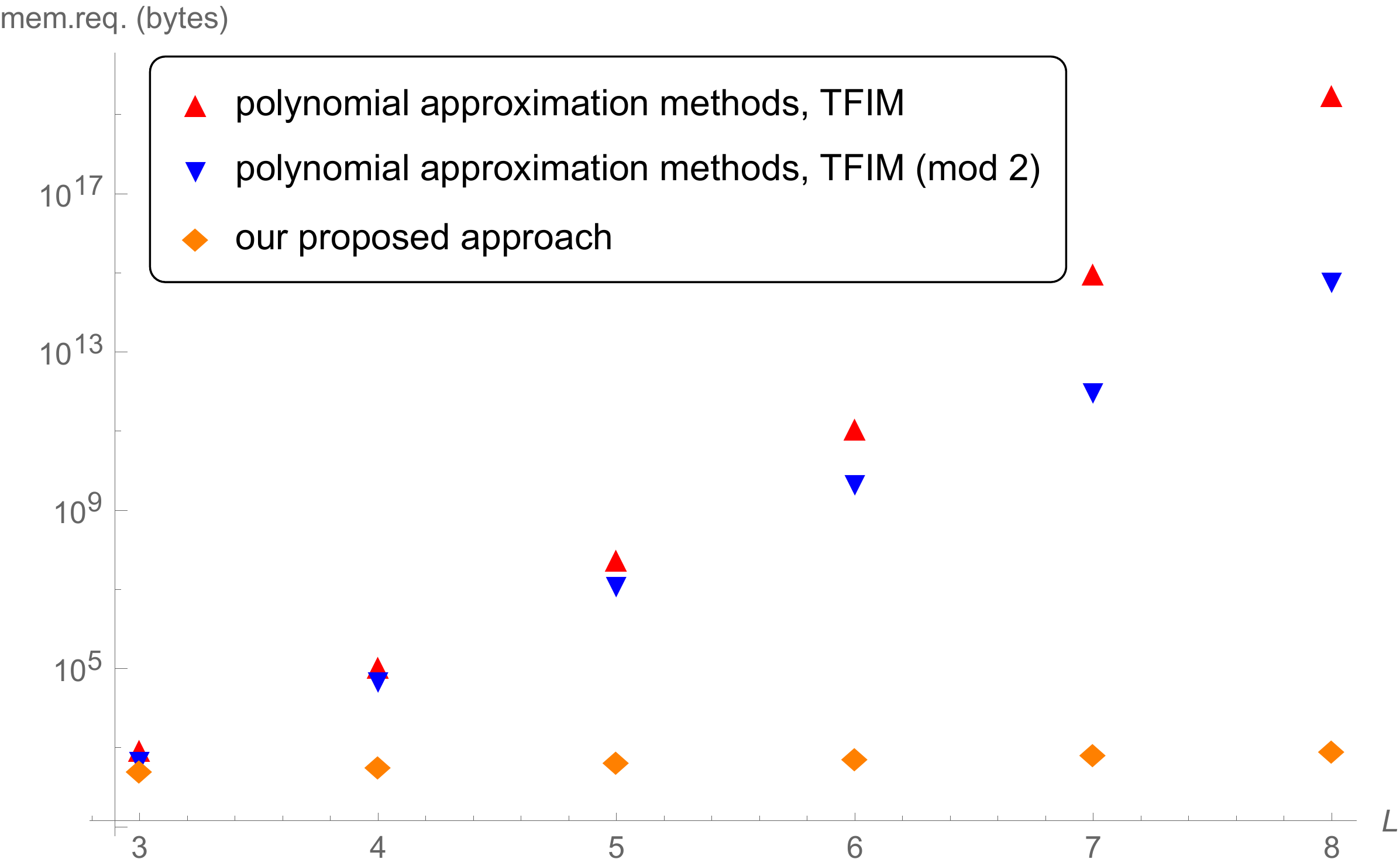}
\caption{\label{fig:memory}
Memory requirements for calculating an element of the matrix exponential
for the TFIM(mod 2) and TFIM matrices. For each of these two cases, 
polynomial expansion-based and Krylov subspace-based methods 
require storing at least one vector with $A_q$ nonzero elements,
where $q$ is taken from Table~\ref{tab:degree} 
and the lower bounds for $A_q$ are indicated in Sec.~\ref{sub:poly}.
The proposed approach needs only about $3n+5Q$ variables for each of the cases TFIM(mod 2) and TFIM.}
\end{figure}

In order to substantially reduce the numerical rounding error of the summation in Eq.~(\ref{eq:matexp}),
a compensated summation algorithm can be used~\cite{kahan1965,higham1993}.
The most well-known compensated summation is the Kahan summation algorithm, where
the numerical error $E_N=\hat S_N - S_N$ of the sum $S_N=\sum_{i=1}^N x_i$ is known to be bounded as follows~\cite{higham1993}:
\beq
\frac{|E_N|}{|S_N|}\leq [2\varepsilon+O(N\varepsilon^2)] \frac{\sum_{i=1}^N |x_i|}{\left|\sum_{i=1}^N x_i\right|},
\label{eq:kahanbound}
\eeq
where $\varepsilon$ is the machine precision of the arithmetic being used, i.e., $\varepsilon\approx 10^{-16}$ for  double-precision
floating point and $\varepsilon\approx 10^{-19}$ for  80-bit extended-precision floating point.
Since all summands in Eq.~(\ref{eq:matexp}) for a given $q$ share the same sign, the relative error
is bounded by $2\varepsilon+O(N\varepsilon^2)$ and is effectively independent of $N$ for $N \leq 1/\varepsilon$.
The bound, Eq.~(\ref{eq:kahanbound}), is substantially better than the worst-case error
of the recursive (naive) summation algorithm, where $|E_N|/|S_N|\leq (N-1)\varepsilon\sum_{i=1}^N |x_i|/\left|S_N\right|$.

We plot in Fig.~\ref{fig:error} the estimated relative error 
of a diagonal matrix element calculation as a function of the 
off-diagonal strength $\Gamma$ and maximal expansion order $Q$, where TFIM(mod 2) and TFIM denote
the algorithms described in Secs.~\ref{sec:isingmod2} and~\ref{sec:ising}, respectively.
Here, the basis vector was chosen randomly. It follows that for $Q=8$ a reasonable accuracy can be achieved 
for $\Gamma \leq 0.05$ and $\Gamma \leq 0.01$ for TFIM(mod 2) and TFIM, respectively.
Figure~\ref{fig:error:nondiag} shows the estimated relative error for a nondiagonal matrix element calculation.
Figure~\ref{fig:time} shows the estimated wall-clock time of a sequential program code as a function of $Q$ for
the TFIM(mod 2) and TFIM algorithms. The sequential wall-clock time grows very quickly with $Q$.

We have additionally verified the correctness and accuracy of the computation by ensuring that the calculated values 
exactly coincide with the calculation results by other methods for $L=3$ and $L=4$, where the matrix size is substantially 
smaller and so that other methods can be applied.

As already discussed above, a key feature of our algorithm is that it is memory efficient.
In Fig.~\ref{fig:memory} we illustrate that point. For the TFIM(mod 2) and TFIM matrices, 
the proposed method requires an exponentially smaller amount of memory
for calculating a matrix element of the matrix exponential
compared to polynomial expansion-based and Krylov subspace-based methods.
The inefficiency of the polynomial expansion-based methods
for these cases is a consequence of the lower bounds for $A_q$ in Sec.~\ref{sub:poly}.

In addition, we note that the proposed algorithm can be efficiently parallelized.
One way to do so is to use a distinct parallel thread for each set of numbers
$k_1,\dots,k_n \geq 0$ such that $k_1+\dots+k_n=q$, $k_1,\dots,k_m$ are odd, and
$k_{m+1},\dots,k_n$ are even, where $m$ is number of different bits between $i$ and $f$.
Each thread generates all corresponding walks and computes respective divided differences
as described in the second part of Sec.~\ref{sec:walkGeneration}.
The number of such sets of numbers $\binom{(q-m)/2+n-1}{n-1}$ is sufficiently large
when $q>m$, which is true in most cases. If this number is small, one can use the alternative method and perform the routine in the second part of Sec.~\ref{sec:walkGeneration} using many parallel threads. This can be done by separating the first layer or several layers of the recursion such that each parallel thread computes the remaining layers. In this case, each thread corresponds to a sequence of $X_j$ operators of a given small length, and it generates all walks of length $q$ whose initial part coincides with the given one.

\subsection{Calculating quantum transition amplitudes}

Another matrix function that is of major practical interest in physics is $f(x)=\e^{-i t x}$ for real $t$. 
The elements of $f(M)$, namely $\langle z_\fin| f(M)|z_\ini\rangle$, in the case where $M$ is taken to be the Hamiltonian of a given physical system, correspond to quantum transition amplitudes, the norms of which represent the probabilities of transitioning from the initial state $|z_\ini\rangle$ to the final state $|z_\fin\rangle$ under the dynamics generated by $M$. 

For the above function, each element can be expressed as Eq.~(\ref{eq:matexp}),
where $\beta=i t$, so the calculation can be performed in the same way
as described in Sec.~\ref{sec:ising}.
The difference is that the input lists of the divided differences now contain
complex numbers, and the divided differences themselves are complex numbers as well.
The algorithms described in Refs.~\cite{divdiff2020,Zivcovich2019},
which calculate the divided differences, are applicable to the case
of a list of complex numbers. Therefore, the algorithms
described above can be directly applied for computing the quantum transition amplitudes
for the Hamiltonians~ Eqs.~(\ref{hamiltonian}) and ~(\ref{hamiltonianmod2}).
In particular, the computational complexity of calculating $\langle z_\fin |\e^{-i t M} | z_\ini \rangle$ 
is close to that of $\langle z_\fin |\e^{-|t| M} | z_\ini \rangle$.

\begin{figure}[tb]
\includegraphics[width=0.6\textwidth]{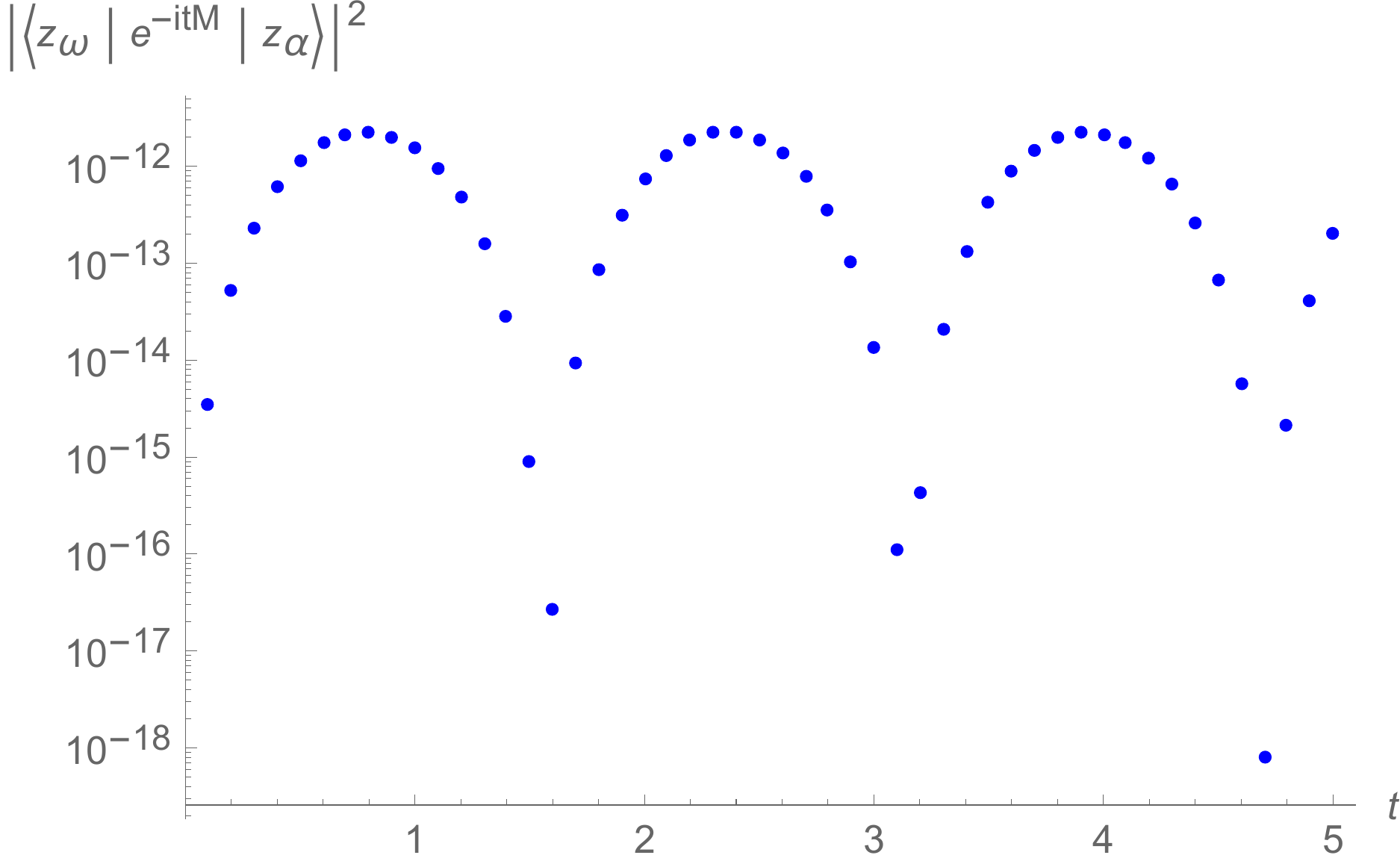}
\caption{\label{fig:tfimamplitudes} Transition probabilities 
$\left|\langle z_\fin | \exp(-i t M)  | z_\ini \rangle\right|^2$, where $M$ is the transverse-field Ising model~\eqref{hamiltonian} with
$L=8$, $\Gamma=0.001$, $J_{ij}=1$. Here, $\ini=16210525687446977967$, $\fin=16489678495598722447$,
so the number of different bits between $\ini$ and $\fin$ is $m=5$.}
\end{figure}

As an example of such calculation, Fig.~\ref{fig:tfimamplitudes} shows transition probabilities
$\left|\langle z_\fin | \exp(-i t M) | z_\ini \rangle\right|^2$ as a function of $t$, where $M$ is the 
transverse-field Ising model, Eq.~\eqref{hamiltonian} with $L=8$, $\Gamma=0.001$, $J_{ij}=1$,
$z_\ini$ is a randomly chosen basis vector, and $z_\fin$ is a basis vector
such that the number of different bits between $\ini$ and $\fin$ is $m=5$.

\section{Solving a dense linear system of equations\label{sec:denseLin}}

We next consider the problem of calculating matrix elements of the inverse 
of a \emph{dense} matrix $M$. 
Of course, the size of dense matrices for which the calculations can be performed 
is much smaller than the size of the sparse matrices discussed in the previous section.

For $f(z)=z^{-1}$ we have an explicit form of the divided differences as
\beq
\label{ddinv}
f[z_0,\ldots,z_q] = \frac{(-1)^q}{\prod_{k=0}^{q} z_k} \,.
\eeq

The expression~(\ref{ddinv}) is ill-defined if any of the inputs is zero, so
for simplicity, let us assume that the diagonal elements of matrix $M$ are nonzero.
One can get rid of the diagonal terms setting them all to $1$ by applying a transformation.
Indeed, the linear system of equations
$M |x\rangle = |b\rangle$ is equivalent to the system of equations
$\tilde M |\tilde x\rangle = |\tilde b\rangle$, where 
$\tilde M = DMD$, $|\tilde x\rangle = D^{-1} |x\rangle$, $|\tilde b\rangle = D|b\rangle$,
and $D = \mathrm{diag}\{m_{ii}^{-1/2}\}$.
Then, all diagonal values of the matrix $\tilde M$ are equal to $1$.
Applying the transformation speeds up the calculation of the sum Eq.~(\ref{eq:matEl}),
which then results in solving the system of equations with only $O(q N^2)$ flops and
$O(N)$ bytes of memory, where $N$ is the matrix size.

When all diagonal values of the matrix $M$ are equal to $1$, Eq.~(\ref{eq:matEl}) can be rewritten as
\beq
\langle z_{\fin} | M^{-1}|z_{\ini}\rangle =
\sum_{q=0}^{\infty} \sum_{\langle z_{\fin} | S_{{\bf{i}}_q} |z_{\ini}\rangle=1} 
(-1)^q \langle z_q | M | z_{q-1} \rangle\langle z_{q-1} | M | z_{q-2} \rangle\dots\langle z_1 | M | z_0 \rangle =
\sum_{q=0}^{\infty} (-1)^q \langle z_{\fin} | T^q | z_{\ini}\rangle\,,
\eeq
where $T = M - \openone$.
Therefore, our proposed method reduces in this case to simply evaluating the Neumann series
$$
M^{-1}|z\rangle = (\openone+T)^{-1}|z\rangle = \sum_{q=0}^{\infty} (-1)^q\, T^q | z\rangle\,.
$$
The spectral radius of $T$ needs to be smaller than $1$ for this series to converge.

\section{Summary and outlook\label{sec:conclusions}}
 
We developed a memory efficient method for calculating the elements 
of functions of extremely large matrices using the off-diagonal series expansion.
In our approach, each matrix element is expressed as a sum of terms 
each of which is a divided difference corresponding to a walk on the graph defined by the matrix in question.
We demonstrated that our method is applicable to very large matrices as long as the 
off-diagonal elements are sufficiently small. We also showed that the algorithm
is highly parallelizable.

The detailed algorithms for calculating individual matrix elements
have been described for several physically motivated examples describing large many-body Hamiltonians
such as exponential of a transverse-field Ising model. To showcase the applicability and scope of our method, we have calculated matrix elements of the exponentials of matrices of sizes up to $2^{64} \times 2^{64}$ on a single workstation. Such calculations are allowed due to the fact that the memory requirements do not grow with the dimension of the matrix as is the case with existing methods. 

In addition to matrix exponentials, we also considered the calculation of matrix elements of the inverse of a dense matrix and showed that in this case the method reduces to the evaluation of the Neumann series and thus does not provide any advantages.
One of the interesting questions which deserves further investigation is the possibility of further exact resummations of walks in the framework of divided differences such as reformulation of path-sums of~\cite{giscard2013evaluating}.

We have demonstrated that the method can be applicable and is efficient in many cases for which existing polynomial approximation methods are not. We therefore hope that our method becomes a useful, even indispensable, tool in areas where matrix functions are needed, such as calculating quantum transition amplitudes, quantum partition functions and beyond.

\begin{acknowledgments}
LB acknowledges support within the framework of State Assignment No. 0029-2019-0003 of Russian Ministry of Science and Higher Education. SG by the UK's Alan Turing Institute under the EPSRC grant EP/N510129/1. 
Work by IH was supported by the U.S. Department of Energy (DOE), Office of Science, Basic Energy Sciences (BES) under Award No. DE-SC0020280.
\end{acknowledgments}
\bibliography{refs}

\begin{thebibliography}{25}%
\makeatletter
\providecommand \@ifxundefined [1]{%
 \@ifx{#1\undefined}
}%
\providecommand \@ifnum [1]{%
 \ifnum #1\expandafter \@firstoftwo
 \else \expandafter \@secondoftwo
 \fi
}%
\providecommand \@ifx [1]{%
 \ifx #1\expandafter \@firstoftwo
 \else \expandafter \@secondoftwo
 \fi
}%
\providecommand \natexlab [1]{#1}%
\providecommand \enquote  [1]{``#1''}%
\providecommand \bibnamefont  [1]{#1}%
\providecommand \bibfnamefont [1]{#1}%
\providecommand \citenamefont [1]{#1}%
\providecommand \href@noop [0]{\@secondoftwo}%
\providecommand \href [0]{\begingroup \@sanitize@url \@href}%
\providecommand \@href[1]{\@@startlink{#1}\@@href}%
\providecommand \@@href[1]{\endgroup#1\@@endlink}%
\providecommand \@sanitize@url [0]{\catcode `\\12\catcode `\$12\catcode
  `\&12\catcode `\#12\catcode `\^12\catcode `\_12\catcode `\%12\relax}%
\providecommand \@@startlink[1]{}%
\providecommand \@@endlink[0]{}%
\providecommand \url  [0]{\begingroup\@sanitize@url \@url }%
\providecommand \@url [1]{\endgroup\@href {#1}{\urlprefix }}%
\providecommand \urlprefix  [0]{URL }%
\providecommand \Eprint [0]{\href }%
\providecommand \doibase [0]{https://doi.org/}%
\providecommand \selectlanguage [0]{\@gobble}%
\providecommand \bibinfo  [0]{\@secondoftwo}%
\providecommand \bibfield  [0]{\@secondoftwo}%
\providecommand \translation [1]{[#1]}%
\providecommand \BibitemOpen [0]{}%
\providecommand \bibitemStop [0]{}%
\providecommand \bibitemNoStop [0]{.\EOS\space}%
\providecommand \EOS [0]{\spacefactor3000\relax}%
\providecommand \BibitemShut  [1]{\csname bibitem#1\endcsname}%
\let\auto@bib@innerbib\@empty
%</preamble>
\bibitem [{\citenamefont {Higham}(2008)}]{fom}%
  \BibitemOpen
  \bibfield  {author} {\bibinfo {author} {\bibfnamefont {J.~N.}\ \bibnamefont
  {Higham}},\ }\href {https://doi.org/10.1137/1.9780898717778} {\emph {\bibinfo
  {title} {Functions of matrices: Theory and Computation}}}\ (\bibinfo
  {publisher} {Society for Industrial and Applied Mathematics},\ \bibinfo
  {year} {2008})\BibitemShut {NoStop}%
\bibitem [{\citenamefont {G\"{u}ttel}\ \emph {et~al.}(2020)\citenamefont
  {G\"{u}ttel}, \citenamefont {Kressner},\ and\ \citenamefont {Lund}}]{GKL20}%
  \BibitemOpen
  \bibfield  {author} {\bibinfo {author} {\bibfnamefont {S.}~\bibnamefont
  {G\"{u}ttel}}, \bibinfo {author} {\bibfnamefont {D.}~\bibnamefont
  {Kressner}},\ and\ \bibinfo {author} {\bibfnamefont {K.}~\bibnamefont
  {Lund}},\ }\href {https://doi.org/10.1002/gamm.202000019} {\bibfield
  {journal} {\bibinfo  {journal} {GAMM-Mitteilungen}\ }\textbf {\bibinfo
  {volume} {43}},\ \bibinfo {pages} {e202000019} (\bibinfo {year}
  {2020})}\BibitemShut {NoStop}%
\bibitem [{\citenamefont {Higham}\ and\ \citenamefont
  {Hopkins}(2020)}]{higham2020catalogue}%
  \BibitemOpen
  \bibfield  {author} {\bibinfo {author} {\bibfnamefont {N.~J.}\ \bibnamefont
  {Higham}}\ and\ \bibinfo {author} {\bibfnamefont {E.}~\bibnamefont
  {Hopkins}},\ }\href {http://eprints.maths.manchester.ac.uk/2754/} {\emph
  {\bibinfo {title} {{A Catalogue of Software for Matrix Functions. Version
  3.0}}}},\ \bibinfo {type} {MIMS EPrint}\ \bibinfo {number} {2020.7}\
  (\bibinfo  {institution} {Manchester Institute for Mathematical Sciences, The
  University of Manchester},\ \bibinfo {address} {UK},\ \bibinfo {year}
  {2020})\BibitemShut {NoStop}%
\bibitem [{\citenamefont {G{\"u}ttel}(2013)}]{Gue13b}%
  \BibitemOpen
  \bibfield  {author} {\bibinfo {author} {\bibfnamefont {S.}~\bibnamefont
  {G{\"u}ttel}},\ }\href {https://doi.org/10.1002/gamm.201310002} {\bibfield
  {journal} {\bibinfo  {journal} {GAMM-Mitteilungen}\ }\textbf {\bibinfo
  {volume} {36}},\ \bibinfo {pages} {8} (\bibinfo {year} {2013})}\BibitemShut
  {NoStop}%
\bibitem [{\citenamefont {Albash}\ \emph {et~al.}(2017)\citenamefont {Albash},
  \citenamefont {Wagenbreth},\ and\ \citenamefont {Hen}}]{ODE}%
  \BibitemOpen
  \bibfield  {author} {\bibinfo {author} {\bibfnamefont {T.}~\bibnamefont
  {Albash}}, \bibinfo {author} {\bibfnamefont {G.}~\bibnamefont {Wagenbreth}},\
  and\ \bibinfo {author} {\bibfnamefont {I.}~\bibnamefont {Hen}},\ }\href
  {https://doi.org/10.1103/PhysRevE.96.063309} {\bibfield  {journal} {\bibinfo
  {journal} {Phys. Rev. E}\ }\textbf {\bibinfo {volume} {96}},\ \bibinfo
  {pages} {063309} (\bibinfo {year} {2017})}\BibitemShut {NoStop}%
\bibitem [{\citenamefont {Hen}(2018)}]{ODE2}%
  \BibitemOpen
  \bibfield  {author} {\bibinfo {author} {\bibfnamefont {I.}~\bibnamefont
  {Hen}},\ }\href {https://doi.org/10.1088/1742-5468/aabbe4} {\bibfield
  {journal} {\bibinfo  {journal} {Journal of Statistical Mechanics: Theory and
  Experiment}\ }\textbf {\bibinfo {volume} {2018}},\ \bibinfo {pages} {053102}
  (\bibinfo {year} {2018})}\BibitemShut {NoStop}%
\bibitem [{\citenamefont {Gupta}\ \emph
  {et~al.}(2020{\natexlab{a}})\citenamefont {Gupta}, \citenamefont {Albash},\
  and\ \citenamefont {Hen}}]{pmr}%
  \BibitemOpen
  \bibfield  {author} {\bibinfo {author} {\bibfnamefont {L.}~\bibnamefont
  {Gupta}}, \bibinfo {author} {\bibfnamefont {T.}~\bibnamefont {Albash}},\ and\
  \bibinfo {author} {\bibfnamefont {I.}~\bibnamefont {Hen}},\ }\href
  {https://doi.org/10.1088/1742-5468/ab9e64} {\bibfield  {journal} {\bibinfo
  {journal} {Journal of Statistical Mechanics: Theory and Experiment}\ }\textbf
  {\bibinfo {volume} {2020}},\ \bibinfo {pages} {073105} (\bibinfo {year}
  {2020}{\natexlab{a}})}\BibitemShut {NoStop}%
\bibitem [{\citenamefont {Cvetkovi{\'c}}\ \emph {et~al.}(1997)\citenamefont
  {Cvetkovi{\'c}}, \citenamefont {Rowlinson},\ and\ \citenamefont
  {Simic}}]{cvetkovic1997eigenspaces}%
  \BibitemOpen
  \bibfield  {author} {\bibinfo {author} {\bibfnamefont {D.~M.}\ \bibnamefont
  {Cvetkovi{\'c}}}, \bibinfo {author} {\bibfnamefont {P.}~\bibnamefont
  {Rowlinson}},\ and\ \bibinfo {author} {\bibfnamefont {S.}~\bibnamefont
  {Simic}},\ }\href {https://doi.org/10.1017/CBO9781139086547} {\emph {\bibinfo
  {title} {Eigenspaces of Graphs}}},\ \bibinfo {number} {66}\ (\bibinfo
  {publisher} {Cambridge University Press},\ \bibinfo {year}
  {1997})\BibitemShut {NoStop}%
\bibitem [{\citenamefont {Estrada}\ and\ \citenamefont
  {Higham}(2010)}]{estrada2010network}%
  \BibitemOpen
  \bibfield  {author} {\bibinfo {author} {\bibfnamefont {E.}~\bibnamefont
  {Estrada}}\ and\ \bibinfo {author} {\bibfnamefont {D.~J.}\ \bibnamefont
  {Higham}},\ }\href {https://doi.org/10.1137/090761070} {\bibfield  {journal}
  {\bibinfo  {journal} {SIAM Review}\ }\textbf {\bibinfo {volume} {52}},\
  \bibinfo {pages} {696} (\bibinfo {year} {2010})}\BibitemShut {NoStop}%
\bibitem [{\citenamefont {Giscard}\ \emph {et~al.}(2013)\citenamefont
  {Giscard}, \citenamefont {Thwaite},\ and\ \citenamefont
  {Jaksch}}]{giscard2013evaluating}%
  \BibitemOpen
  \bibfield  {author} {\bibinfo {author} {\bibfnamefont {P.-L.}\ \bibnamefont
  {Giscard}}, \bibinfo {author} {\bibfnamefont {S.~J.}\ \bibnamefont
  {Thwaite}},\ and\ \bibinfo {author} {\bibfnamefont {D.}~\bibnamefont
  {Jaksch}},\ }\href {https://doi.org/10.1137/120862880} {\bibfield  {journal}
  {\bibinfo  {journal} {SIAM Journal on Matrix Analysis and Applications}\
  }\textbf {\bibinfo {volume} {34}},\ \bibinfo {pages} {445} (\bibinfo {year}
  {2013})}\BibitemShut {NoStop}%
\bibitem [{\citenamefont {Joyner}(2008)}]{gpm}%
  \BibitemOpen
  \bibfield  {author} {\bibinfo {author} {\bibfnamefont {D.}~\bibnamefont
  {Joyner}},\ }\href@noop {} {\emph {\bibinfo {title} {Adventures in group
  theory. Rubik's cube, Merlin's machine, and other mathematical toys}}}\
  (\bibinfo  {publisher} {Baltimore, MD: Johns Hopkins University Press},\
  \bibinfo {year} {2008})\BibitemShut {NoStop}%
\bibitem [{\citenamefont {Hen}(2019)}]{signProbODE}%
  \BibitemOpen
  \bibfield  {author} {\bibinfo {author} {\bibfnamefont {I.}~\bibnamefont
  {Hen}},\ }\href {https://doi.org/10.1103/PhysRevE.99.033306} {\bibfield
  {journal} {\bibinfo  {journal} {Phys. Rev. E}\ }\textbf {\bibinfo {volume}
  {99}},\ \bibinfo {pages} {033306} (\bibinfo {year} {2019})}\BibitemShut
  {NoStop}%
\bibitem [{\citenamefont {Whittaker}\ and\ \citenamefont
  {Robinson}(1940)}]{dd:67}%
  \BibitemOpen
  \bibfield  {author} {\bibinfo {author} {\bibfnamefont {E.~T.}\ \bibnamefont
  {Whittaker}}\ and\ \bibinfo {author} {\bibfnamefont {G.}~\bibnamefont
  {Robinson}},\ }\href@noop {} {\emph {\bibinfo {title} {The Calculus of
  Observations: A Treatise on Numerical Mathematics, 3rd Edition.}}}\ (\bibinfo
   {publisher} {Blackie and Sons Limited, London},\ \bibinfo {year}
  {1940})\BibitemShut {NoStop}%
\bibitem [{\citenamefont {de~Boor}(2005)}]{deboor:05}%
  \BibitemOpen
  \bibfield  {author} {\bibinfo {author} {\bibfnamefont {C.}~\bibnamefont
  {de~Boor}},\ }\href@noop {} {\bibfield  {journal} {\bibinfo  {journal}
  {Surveys in Approximation Theory}\ }\textbf {\bibinfo {volume} {1}},\
  \bibinfo {pages} {46} (\bibinfo {year} {2005})}\BibitemShut {NoStop}%
\bibitem [{\citenamefont {Gupta}\ \emph
  {et~al.}(2020{\natexlab{b}})\citenamefont {Gupta}, \citenamefont {Barash},\
  and\ \citenamefont {Hen}}]{divdiff2020}%
  \BibitemOpen
  \bibfield  {author} {\bibinfo {author} {\bibfnamefont {L.}~\bibnamefont
  {Gupta}}, \bibinfo {author} {\bibfnamefont {L.}~\bibnamefont {Barash}},\ and\
  \bibinfo {author} {\bibfnamefont {I.}~\bibnamefont {Hen}},\ }\href
  {https://doi.org/10.1016/j.cpc.2020.107385} {\bibfield  {journal} {\bibinfo
  {journal} {Computer Physics Communications}\ }\textbf {\bibinfo {volume}
  {254}},\ \bibinfo {pages} {107385} (\bibinfo {year}
  {2020}{\natexlab{b}})}\BibitemShut {NoStop}%
\bibitem [{\citenamefont {Pfeuty}\ and\ \citenamefont {Elliott}(1971)}]{tfim1}%
  \BibitemOpen
  \bibfield  {author} {\bibinfo {author} {\bibfnamefont {P.}~\bibnamefont
  {Pfeuty}}\ and\ \bibinfo {author} {\bibfnamefont {R.~J.}\ \bibnamefont
  {Elliott}},\ }\href {https://doi.org/10.1088/0022-3719/4/15/024} {\bibfield
  {journal} {\bibinfo  {journal} {Journal of Physics C: Solid State Physics}\
  }\textbf {\bibinfo {volume} {4}},\ \bibinfo {pages} {2370} (\bibinfo {year}
  {1971})}\BibitemShut {NoStop}%
\bibitem [{\citenamefont {Stinchcombe}(1973)}]{tfim2}%
  \BibitemOpen
  \bibfield  {author} {\bibinfo {author} {\bibfnamefont {R.~B.}\ \bibnamefont
  {Stinchcombe}},\ }\href {https://doi.org/10.1088/0022-3719/6/15/009}
  {\bibfield  {journal} {\bibinfo  {journal} {Journal of Physics C: Solid State
  Physics}\ }\textbf {\bibinfo {volume} {6}},\ \bibinfo {pages} {2459}
  (\bibinfo {year} {1973})}\BibitemShut {NoStop}%
\bibitem [{Note1()}]{Note1}%
  \BibitemOpen
  \bibinfo {note} {If $L$ is odd, the first summand in~(\ref {hamiltonianmod2})
  should be understood as $\left \lfloor \protect \frac 14 \left | \DOTSB \sum@
  \slimits@ _{\langle i,j\rangle } Z_i Z_j \right | \right \rfloor \ (\protect
  \mathrm {mod}\ 2)$, where $\lfloor x\rfloor $ denotes the largest integer
  that is less than or equal to $x$.}\BibitemShut {Stop}%
\bibitem [{Mat()}]{MatrixFunctionsCode}%
  \BibitemOpen
  \href@noop {} {\bibinfo {title} {Matrix functions program codes in c++}},\
  \bibinfo {howpublished}
  {\url{https://github.com/LevBarash/MatrixFunctions}}\BibitemShut {NoStop}%
\bibitem [{\citenamefont {Zivcovich}(2019)}]{Zivcovich2019}%
  \BibitemOpen
  \bibfield  {author} {\bibinfo {author} {\bibfnamefont {F.}~\bibnamefont
  {Zivcovich}},\ }\href {https://doi.org/10.14658/pupj-drna-2019-1-4}
  {\bibfield  {journal} {\bibinfo  {journal} {Dolomites Research Notes on
  Approximation}\ }\textbf {\bibinfo {volume} {12}},\ \bibinfo {pages} {28}
  (\bibinfo {year} {2019})}\BibitemShut {NoStop}%
\bibitem [{\citenamefont {Farwig}\ and\ \citenamefont
  {Zwick}(1985)}]{Farwig1985}%
  \BibitemOpen
  \bibfield  {author} {\bibinfo {author} {\bibfnamefont {R.}~\bibnamefont
  {Farwig}}\ and\ \bibinfo {author} {\bibfnamefont {D.}~\bibnamefont {Zwick}},\
  }\href {https://doi.org/10.1016/0022-247X(85)90036-8} {\bibfield  {journal}
  {\bibinfo  {journal} {Journal of Mathematical Analysis and Applications}\
  }\textbf {\bibinfo {volume} {108}},\ \bibinfo {pages} {430} (\bibinfo {year}
  {1985})}\BibitemShut {NoStop}%
\bibitem [{\citenamefont {Druskin}\ and\ \citenamefont
  {Knizhnerman}(1989)}]{druskin1989two}%
  \BibitemOpen
  \bibfield  {author} {\bibinfo {author} {\bibfnamefont {V.~L.}\ \bibnamefont
  {Druskin}}\ and\ \bibinfo {author} {\bibfnamefont {L.~A.}\ \bibnamefont
  {Knizhnerman}},\ }\href {https://doi.org/10.1016/S0041-5553(89)80020-5}
  {\bibfield  {journal} {\bibinfo  {journal} {USSR Computational Mathematics
  and Mathematical Physics}\ }\textbf {\bibinfo {volume} {29}},\ \bibinfo
  {pages} {112} (\bibinfo {year} {1989})}\BibitemShut {NoStop}%
\bibitem [{\citenamefont {Driscoll}\ \emph {et~al.}(2014)\citenamefont
  {Driscoll}, \citenamefont {Hale},\ and\ \citenamefont
  {Trefethen}}]{driscoll2014chebfun}%
  \BibitemOpen
  \bibfield  {author} {\bibinfo {author} {\bibfnamefont {T.~A.}\ \bibnamefont
  {Driscoll}}, \bibinfo {author} {\bibfnamefont {N.}~\bibnamefont {Hale}},\
  and\ \bibinfo {author} {\bibfnamefont {L.~N.}\ \bibnamefont {Trefethen}},\
  }\href {https://www.chebfun.org/docs/guide/} {\bibinfo {title} {Chebfun
  guide}} (\bibinfo {year} {2014})\BibitemShut {NoStop}%
\bibitem [{\citenamefont {Kahan}(1965)}]{kahan1965}%
  \BibitemOpen
  \bibfield  {author} {\bibinfo {author} {\bibfnamefont {W.}~\bibnamefont
  {Kahan}},\ }\href {https://doi.org/10.1145/363707.363723} {\bibfield
  {journal} {\bibinfo  {journal} {Commun. ACM}\ }\textbf {\bibinfo {volume}
  {8}},\ \bibinfo {pages} {40} (\bibinfo {year} {1965})}\BibitemShut {NoStop}%
\bibitem [{\citenamefont {Higham}(1993)}]{higham1993}%
  \BibitemOpen
  \bibfield  {author} {\bibinfo {author} {\bibfnamefont {N.~J.}\ \bibnamefont
  {Higham}},\ }\href {https://doi.org/10.1137/0914050} {\bibfield  {journal}
  {\bibinfo  {journal} {SIAM Journal on Scientific Computing}\ }\textbf
  {\bibinfo {volume} {14}},\ \bibinfo {pages} {783} (\bibinfo {year}
  {1993})}\BibitemShut {NoStop}%
\end{thebibliography}%

\begin{appendix}

\section{Notes on divided differences\label{app:dd}}

We provide below a brief summary of the concept of divided differences, which is a recursive division process. This method is typically encountered when calculating the coefficients in the interpolation polynomial in the Newton form.

The divided differences~\cite{dd:67,deboor:05} of a function $f(\cdot)$ are defined as
\beq\label{eq:divideddifference2}
f[x_0,\ldots,x_q] \equiv \sum_{j=0}^{q} \frac{f(x_j)}{\prod_{k \neq j}(x_j-x_k)}
\eeq
with respect to the list of real-valued input variables $[x_0,\ldots,x_q]$. The above expression is ill-defined if some of the inputs have repeated values, in which case one must resort to the use of limits. For instance, in the case where $x_0=x_1=\ldots=x_q=x$, the definition of divided differences reduces to: 
\beq
f[x_0,\ldots,x_q] = \frac{f^{(q)}(x)}{q!} \,,
\eeq 
where $f^{(n)}(\cdot)$ stands for the $n$-th derivative of $f(\cdot)$.
Divided differences can alternatively be defined via the recursion relations
\bea\label{eq:ddr}
f[x_i,\ldots,x_{i+j}] = \frac{f[x_{i+1},\ldots , x_{i+j}] - f[x_i,\ldots , x_{i+j-1}]}{x_{i+j}-x_i} \,,
\eea 
with $i\in\{0,\ldots,q-j\},\ j\in\{1,\ldots,q\}$ and the initial conditions
\beq\label{eq:divideddifference3}
f[x_i] = f(x_{i}), \qquad i \in \{ 0,\ldots,q \}  \quad \forall i \,.
\eeq
A function of divided differences can be defined in terms of its Taylor expansion
\beq
f[x_0,\ldots,x_q] = \sum_{n=0}^\infty \frac{f^{(n)}(0)}{n!} [x_0,\ldots,x_q]^n \ .
\eeq
Moreover, it is easy to verify that
\beq  \label{eq:ts}
[x_0,\ldots,x_q]^{q+n} = \Bigg\{ 
\begin{tabular}{ l c l }
   $0$ & \phantom{$0$} & $n<0$ \\
  $1$ & \phantom{$0$} &  $n=0$ \\
  $\sum_{\sum k_j = n} \prod _{j=0}^{q} x_j^{k_j}$& \phantom{$0$} &  $n>0$ \\
\end{tabular}
 \,.
\eeq
One may therefore write:
\beq
f[x_0,\ldots,x_q] = \sum_{n=0}^\infty \frac{f^{(n)}(0)}{n!} [x_0,\ldots,x_q]^n =
\sum_{n=q}^\infty \frac{f^{(n)}(0)}{n!} [x_0,\ldots,x_q]^n =
\sum_{m=0}^\infty \frac{f^{(q+m)}(0)}{(q+m)!} [x_0,\ldots,x_q]^{q+m}.
\eeq
The above expression can be further simplified to
\beq
f[x_0,\ldots,x_q] =
\sum_{\{ k_i\}=(0,\ldots,0)}^{(\infty,\ldots,\infty)}
\frac{f^{(q+\sum k_i)}(0)}{(q+\sum k_i)!} \prod _{j=0}^{q} x_j^{k_j},
\eeq
as was asserted in the main text.
\end{appendix}
\end{document}